\documentclass[11pt]{article}

\usepackage{geometry}                
                        
\usepackage{graphicx}
\usepackage{amsmath}
\usepackage{amssymb}
\usepackage{epstopdf}
\newcommand{\be}{\begin{equation}}
\newcommand{\ee}{\end{equation}}
\newcommand{\eq}[1]{Eq.~(\ref{#1})}
\newcommand{\eqs}[2]{Eqs.~(\ref{#1}) and (\ref{#2})}
\newcommand{\fig}[1]{Fig.~\ref{#1}}
\newcommand{\figs}[2]{Figs.~\ref{#1} and \ref{#2}}
\newcommand{\Sec}[1]{Section~\ref{#1}}
\newcommand{\App}[1]{Appendix~\ref{#1}}

\newcommand{\pythia}{\texttt{Pythia}}

\newcommand{\pythiaeightversion}{\texttt{Pythia~8.108}}
\newcommand{\pythiasixversion}{\texttt{Pythia~6.409}}
\newcommand{\fastjetversion}{\texttt{FastJet~2.3.3}}

\newcommand{\TeV}{\mathrm{~TeV}}
\newcommand{\GeV}{\mathrm{~GeV}}
\newcommand{\Br}{\mathrm{Br}}

\begin{document}
\begin{titlepage}

\title{Strategies to Identify Boosted Tops}

\author{Jesse Thaler$^{1,2}$, Lian-Tao Wang$^{3}$ \\ \\
\textit{$^{1}$ \small Berkeley Center for Theoretical Physics,
  University of California, Berkeley, CA 94720} \\ 
\textit{$^{2}$ \small Theoretical Physics Group, Lawrence Berkeley
  National Laboratory, Berkeley, CA 94720}\\ 
\textit{$^{3}$ \small Department of Physics, Princeton University,
  Princeton, NJ 08540} 
}

\date{}
\maketitle
\thispagestyle{empty}

\begin{abstract}
We study techniques for identifying highly boosted top jets, where the subsequent top decay products are not isolated.  For hadronic boosted tops, we consider variables which probe the jet substructure in order to reduce the background from QCD jets with large invariant mass.  Substructure variables related to two-body kinematics are least sensitive to the modeling of parton shower, while those which involve multi-body kinematics may still have discrimination power.  For leptonic boosted tops, we consider variables which characterize the separation between the lepton---although not isolated by conventional criterion---and the hadronic activity in the top jet.  Such variables are useful in reducing the backgrounds both from heavy-flavor jets and from accidental jet-lepton overlap.  We give numerical estimates of the top identification efficiency versus background rejection rate as a functions of cuts on these variables, and find that these variables offer additional useful information above invariant mass alone. 
\end{abstract}


\end{titlepage}

\tableofcontents

\vspace{.3in}

\section{Motivation}
\label{motivation}

With over 10,000 top events at the Tevatron, CDF and D0
have begun to study the detailed properties and couplings of the top
quark \cite{cdf-top,d0-top}.  At the upcoming Large Hadron Collider
(LHC), the pair 
production cross section for top quarks approaches 1 nb
\cite{stt_lhc}, so with an estimated  100 million top events per year
at peak luminosity, 
ATLAS and CMS will be able to pursue a program of precision top
physics \cite{atlasTDR,cmsTDR}. 

Given the large top Yukawa coupling, it is natural to expect physics
beyond the standard model (BSM) to have strong couplings to the top
sector.  Most solutions to the hierarchy problem require some kind of
top partner
\cite{Dimopoulos:1981zb,UED,ArkaniHamed:2001nc,twinhiggs},
and if that top partner is kinematically 
assessable at the LHC, then its decay products are expected to contain
top quarks \cite{tprime-model,tprime-study-1,tprime-study-2,Nojiri}.
Similarly, if the top quark is a semi-composite 
state \cite{top-composite}, then new resonances in the composite
sector are likely 
to have large branching fractions to top quarks \cite{resonance-ttbar}. 

At first glance, the near 100\% branching ratio $t \rightarrow bW$
implies that any new physics that creates top quarks will be seen in
events with multiple heavy-flavor-tagged jets and leptons.  These
exciting event topologies are somewhat challenging to reconstruct
unambiguously, but in principle one can use standard methods
\cite{top_convention,atlasTDR,cmsTDR} to
identify top quarks coming from new physics.  Indeed, tops produced
nearly at rest in the lab frame have three isolated decay products
which, apart from neutrinos and accidental overlap, can be separately
identified.  So assuming one can isolate a top-rich sample with
well-identified decay products, the main challenge would be to
separate a top-rich BSM signal from the SM top production background. 

On the other hand, it is possible for a top-rich BSM signal to not
have isolated decay products.  If the top is produced with a
considerable boost relative to the lab frame, then the three decay
products from the top will end up in the same part of the detector and
possibly be clustered into a single fat jet.  This occurs for example
with multi-TeV resonances that decay to tops \cite{resonance-ttbar}, i.e. $X
\rightarrow t \bar{t}$, where the boost factor $\gamma_t \sim
m_{X}/(2m_t)$ can be large. 

\begin{figure}
\begin{center}
\includegraphics[scale=0.6]{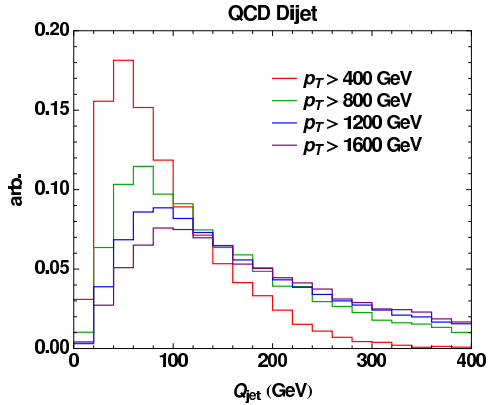}
\end{center}
\caption{\label{fig:jmass-pt} Distribution of the jet invariant mass
  of the hardest 
 jet in QCD dijet production, for $p_T > $ 400, 800, 1200, and 1600
 GeV.  The average jet mass increases with jet $p_T$ roughly as $10 \%
 \times p_T$.  In this and the remaining figures: the QCD 
 production and showering are simulated using \pythia\ \cite{pythia}
 (in this plot \pythiaeightversion); \fastjetversion\ \cite{fastjet}
 is used for jet analysis; jets are identified using the anti-$k_T$ 
 algorithm \cite{anti-kt} with $R_0 = 1.0$; the distributions are
 normalized to 1 to accentuate shape variations. }
\end{figure}

One background to these fat jets comes simply from high-$p_T$ top
events.  Less obvious is that high-$p_T$ QCD events also can yield fat
jets.  As shown in \fig{fig:jmass-pt}, the average invariant mass of a jet is
around 10\% ($\sim \alpha_s$) of its energy,\footnote{Depending on
 exactly what one means by ``average invariant mass'', it may be more
 correct to say $m_{\rm jet} \sim \sqrt{\alpha_s \log (E/E_0)}$,
 where $E_0$ is some reference energy.}  so once QCD jets have 
$p_T \sim 1 \TeV$, the average jet invariant mass approaches the top
mass.  Said another way, the kinematics of top decays and the
kinematics of QCD radiation start to look similar when $\gamma_t \sim
1/\alpha_s$.  Depending on the performance of heavy-flavor tagging at
high-$p_T$, one also worries about fat flavor-tagged jets, which could
come from high-$p_T$ bottom or charm production. 

In this paper, we try to increase the purity for boosted top
identification by finding variables that discriminate between boosted
top jets and ordinary QCD fat jets.  In \Sec{sec:tophadronic}, we
consider boosted hadronic tops and study the energy asymmetry inside
the jet.   Because QCD splitting functions \cite{ap} have soft singularities, it
is likely for a jet to have two subclusters with disparate energy
scales.  In contrast, an on-shell top decay yields a $W$ and a $b$
with roughly comparable energies.  Energy asymmetry turns out to be
the unique fat jet observable one can construct if one approximates
the jet substructure by two-body kinematics, and it offers good
separation between boosted tops from QCD fat jets.  This variable is correlated with the $d_{\rm cut}$ measure used in preliminary boosted top studies at ATLAS \cite{Brooijmans}.   In \Sec{sec:other}, we examine the prospects for identifying top jets
through multi-body signatures.  While the theoretical understanding of
multi-body jet substructure is less controlled, we find that by
identifying the hadronic $W$ within a boosted top, one can obtain
complementary information to energy asymmetry.  Finally, in
\Sec{sec:topleptonic} we look at boosted leptonic tops and study the
kinematics of the muon (or possibly electron) coming from the $W$
decay.  This muon is not isolated when the the top is boosted, but the
muon spectrum does discriminate between a boosted leptonic top, a
bottom jet with a soft muon tag, and accidental jet-lepton overlap. 

The expected effectiveness of these variables depends
crucially on the modeling the high $p_T$ tail of the QCD background.
In the context 
of a Monte Carlo simulation, to build up $170 \GeV$ of radiation in a
jet requires multiple QCD emissions.  While the parton shower
approximation is appropriate for describing emissions in the
soft-collinear limit, the jet mass and the jet substructure depends
crucially on how 
exactly the Monte Carlo program treats the kinematics of successive
emissions. In this study, we focus on variables which are expected to
be least sensitive to the details of the parton showering model, and
we study model variance by using several showering algorithms.
Ultimately, any fat jet  variable will have to be tested on data and
Monte Carlos tuned  appropriately.   

Our study focuses on the differences between QCD fat jets and boosted
top jets without regard to any other features of the event.  In most
cases, the estimated top tagging efficiencies from this paper can be
applied directly on top of a standard event selection criteria.
However, this does mean that we will not discuss the absolute
normalization of the background.  One reason for this choice is that
there are theoretical uncertainties in the size of the QCD jet
invariant mass tail \cite{topmass}, so it is unknown precisely how
much of the QCD dijet background enters the top mass window.  Another
reason is that depending on the specific BSM signal one is interested
in, one would apply additional event selection criteria, and these
criteria would dominantly affect the normalization of background and
only secondarily the spectrum of the hardest fat jet.  On the other
hand, in many interesting new physics processes, boosted tops are pair
produced, so there could be additional useful variables that correlate
the two top quarks.  Indeed, the QCD background to two fat jets will
be different from the background to one fat jet.  A full study of a
boosted $t \bar{t}$ signal is beyond the scope of this paper and will
be left to a future work.

\section{Top Decays vs.\ QCD Radiation}
\label{sec:tophadronic}

\subsection{Jet Substructure Strategy}
\label{subsec:strategy}

When a boosted tops decays hadronically, there are no large sources of
missing energy, so one can identify boosted 
top jets by finding jets with invariant mass $Q^2 \sim m_t^2$. Still,
ordinary QCD jets can have large invariant mass, so one would like to
extract some information about the substructure of the 
jet in order to better distinguish a top decay $t \rightarrow bW$ from
a fat jet formed from QCD radiation.   

\begin{figure}
\begin{center}
\includegraphics[scale=0.6]{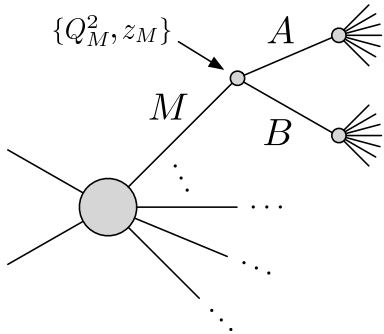}
\end{center}
\caption{\label{fig:2-body} A schematic diagram of a $1 \rightarrow 2$
  process as one stage of a parton shower history.  For our study, the
  mother parton $M$ is assumed to initiate a single jet, and the first
  splitting $M \rightarrow AB$ determines the dominant kinematic
  features of that jet.  After integrating over azimuth, the first
  splitting it is determined by the jet invariant mass $Q_M$ and the
  energy sharing $z_M = \min(E_A,E_B)/E_M$.  Because of a theoretical
  ambiguity as to how to interpret the daughter invariant masses $Q_A$
  and $Q_B$, there are multiple $z_M$ variants, each one giving a
  potential jet substructure observable. 
}
\end{figure}

In the parton shower approximation \cite{reshuffle}, the evolution of
a QCD jet is 
described by a sequence of $1 \rightarrow 2$ splittings, such as the
one shown in Fig.~\ref{fig:2-body}. After averaging over the azimuthal
angle, there are only two variables that characterize the splitting $M
\rightarrow AB$. One 
of them is the invariant mass (or virtuality) of the mother particle
$Q^2_M$.   A typical and convenient choice for the other variable is
the energy sharing 
\be
z_M = \frac{\min(E_A,E_B)}{E_M}.
\ee
All other choices for the two splitting variables are equivalent to
$\{Q^2_M,z_M\}$ up to an ambiguity as to how to treat the invariant
mass of the daughters $A$ and $B$, which we discuss in more detail
below.  For the remainder of this section, we will only look at
variables that can be characterized in terms of an $M \rightarrow AB$
splitting, leaving a discussion of the theoretical challenges and
experimental prospects for multi-body signatures to \Sec{sec:other}. 

Intuitively, differences between a QCD jet and a boosted top jet should
only be significant during the early history of the shower, and the
first splitting should be a ``soft'' splitting in the case of QCD jet
and a``hard'' splitting in the case of the top jet. In Appendix
\ref{app:qcdsoft}, we review two 
theoretical distributions for $z_M$.  The narrow width approximation
is a rough proxy for boosted 
tops, particularly for the first two stages of the prompt decay $t
\rightarrow bW$ and $W \rightarrow q\bar{q}'$.  The QCD parton shower
gives a reasonable approximation to QCD fat jets, especially if we
only look at the first splitting.  In terms of $\{Q^2_M,z_M\}$, the
differential decay distributions for the splitting $M \rightarrow AB$
take roughly the form  
\begin{align}
d f^{\rm NWA}_{M \rightarrow A B} &\sim dQ_M^2 \, d z_M \, \delta(Q_M^2
- m_M^2), \ \ m_M = m_{t} \ \mbox{or} \ m_W  \label{eq:simpsplitnwa} \\ 
d f^{\rm QCD}_{M \rightarrow A B} &\sim dQ_M^2 \, d z_M  \,
\frac{1}{Q_M^2} P(z_M), \label{eq:simpsplitqcd} 
\end{align}
where we have suppressed all non-singular functions.  $P(z_M)$ is an
Alterelli-Parisi splitting function \cite{ap} that typically scales like $1/z_M$.   

Comparing $d f^{\rm NWA}$ and $d f^{\rm QCD}$, the most obvious
difference is the spectrum in $Q_M$, and as we already said, to
distinguish a top jet from a QCD fat jet one first wants isolate a
sample of jets with invariant mass close to $m_t$.   But we see that
the spectrum in energy sharing variable $z_M$ is also different between
top jets and QCD fat jets because of the soft singularity of QCD.  In
particular, $d 
f^{\rm QCD}$ peaks towards $z_M \rightarrow 0$ in QCD, while $d f^{\rm
 NWA}$ has its maximum at some finite value of $z_M$.  If
one can find an experimental observable that is correlated with $z_M$,
then one would have an additional discriminating handle beyond just
jet invariant mass.   

In general, it will be difficult to find an experimental observable
perfectly correlated with a parton shower history.  In order to define
the energy sharing $z_M$, one has to reconstruct the $M \rightarrow
AB$ splitting, but we of course do not actually observe the showering
process.  However, recursive jet clustering algorithms
\cite{jade, durham,geneva,kt,Cambridge-Aachen,anti-kt} are
designed such that the jet clustering procedure approximately reverses
the shower history.  Given a set of particles that forms a jet $M$,
one can apply a clustering algorithm until there are
exactly two subclusters $A$ and $B$.\footnote{It is not necessary to
  use the same algorithm to define $A$ and $B$ as one used to find
  $M$.  In this paper, we will in fact use two different procedures,
  anti-$k_T$ for jet finding and $k_T$ for subcluster finding.} 
As long as the clustering procedure captures the leading
singularities of QCD,\footnote{In order to define $z_M$, it is
  important that the clustering scheme is ``soft-sensitive''.  For
  example, the recently proposed anti-$k_T$ algorithm \cite{anti-kt}
  performs clustering in a way that washes out information about the
  soft-singularity.  Similarly, the distance measure used in the
  Cambridge/Aachen algorithm \cite{Cambridge-Aachen} also does not
  capture the soft 
  singularity of QCD.} the subclusters $A$ and $B$ will be reasonable
proxies for the partons that participated in the first QCD splitting,
and one can define $z_M$ accordingly. 

Of course, even with a perfect clustering algorithm, the
experimentally identified subclusters $A$ and $B$ will 
not have a one-to-one relation to the theoretical splitting $M \rightarrow
AB$.  As an extreme example, even if $A$ and $B$ can be separately
isolated in the rest frame of $M$, if the boost axis of the system is
aligned with the momentum axis of $A$ and $B$, then $A$ and $B$ will
be perfectly overlapping and no experimental observable can
disentangle them.  In addition, finite calorimetry means that the QCD
soft singularity is in effect regulated, so the experimental $z_M$
distribution will always vanish as $z_M \rightarrow 0$.  The hope is
that the experimental and theoretical definitions of $M \rightarrow
AB$ will agree some fraction of the time, so any experimental proxy
for any $z_M$-like variable will offer 
some degree of discrimination power between top jets and QCD fat
jets.   

Finally, a comment about the criteria we use for identifying jets. In
\Sec{subsec:study-z}, we will define jets using a jet radius $R$ which
is larger than the values 
  typically considered by ATLAS or CMS, and one might worry that we
  could obtain more information by using a stricter jet selection
  criteria. 
 For finding boosted tops, though, choosing a smaller value of $R$
 only shifts the problem from 
  tagging single jets with $Q \sim m_t$ to tagging pairs or trios of
  jets with $Q \sim m_t$.  That is, even if on a lego plot it is
  ``obvious'' that a boosted top is composed of three objects, QCD
  fragmentation also has a finite probability of yielding three
  objects with $Q \sim m_t$.  Whether one talks about jet substructure
  on a fat jet or interjet kinematics among smaller jets,
  one is faced with the same problem of trying to determine the
  identity of a set of objects with $Q \sim m_t$.  That said, the
  value of $R$ can certainly be optimized.

\subsection{Choice of $z$ Variable}
\label{subsec:choice-z}

The ideal choice for the $z$ variable is one that is completely
uncorrelated with jet invariant mass for QCD jets.  In that case, one
could imagine doing a sideband analysis to determine the expected QCD
$z$ distribution away from $Q^2 \sim m_t^2$ and then extrapolating to
the top jet region.   In Appendix~\ref{app:qcdsoft}, we argue that in
the leading 
logarithmic approximation, the variable that is most uncorrelated with
invariant mass is the $z_M$ variable that appears in the splitting
function $P(z_M)$.  That is, up to subleading corrections, the precise
functional form of $d f^{\rm QCD}$ in \eq{eq:realsplitqcd} can be
written as the product of a $Q_M^2$-dependent piece and a
$z_M$-dependent piece. 

This pseudo-factorization still does not tell us which definition of
$z_M$ to use.  In order for the Alterelli-Parisi splitting function
$P(z_M)$ to make sense, all that is required is that $z_M  \rightarrow
\min(E_A,E_B) / E_M$ in the limit that $M$, $A$, and $B$ are exactly
collinear and all on-shell.  Given subcluster four vectors $p_A$ and
$p_B$, there are many functions $z_M(p_A,p_B)$ that satisfy this
property, and they differ only by how the virtualities $Q_A^2$ and
$Q_B^2$ of the daughters are treated. 

In the context of a Monte Carlo simulation, each function
$z_M(p_A,p_B)$ corresponds to different assumptions in the 
 parton shower for how to treat successive emissions.  This is
 referred to as momentum reshuffling \cite{reshuffle} and occurs
 because the splitting $M \rightarrow A B$ with on-shell partons $A$
 and $B$ has to be modified when $A$ and $B$ themselves split and go
 off-shell.   The ideal choice for $z$ is one that is uncorrelated
 with $Q_M^2$ \emph{after} taking into account the spectra in $Q_A^2$
 and $Q_B^2$ from subsequent splittings, but
 unfortunately, the information on the correct interpretation of
 $Q_A^2$ and $Q_B^2$ is not contained in \eq{eq:realsplitqcd} and
 requires further insight from QCD.  We discuss this issue more
 in \App{app:qcdsoft}. 
 
Given this ambiguity for how to define $z$, we present three plausible
choices.  The simplest experimental observable correlated with $z$ is  
\be
\label{eq:easyz}
z_{\rm cell} = \frac{\min(E_A,E_B)}{E_A + E_B}, \qquad E_X \equiv
\sum_{i \in X} E_i,  
\ee  
where $E_i$ are the energies of the calorimeter cells and the
subclusters $A$ and $B$ are determined using some clustering scheme.
This definition of $z$ trivially satisfies the property that it
asymptotes to $\min(E_A,E_B) / E_M$ in the exact collinear limit,
though of course in the exact collinear limit, it is impossible to
experimentally separate a jet into subclusters $A$ and
$B$.\footnote{The tension between a theoretical $z$ that has to
 asymptote to a given value in the soft-collinear limit and an
 experimental $z$ that has to know about the $A$/$B$ distinction
 means that $z$ cannot be defined using strictly collinear-safe
 observables.  Said another way, to calculate the QCD $z$ spectrum in
 perturbative theory, one would need to consider \emph{two} partons
 per fat jet.}

While simple, the above definition of $z$ is by no means unique.  A
more complicated choice for a variable correlated with $z$ is the
$d_{\rm cut}$ variable used in the $k_T$ jet clustering algorithm \cite{kt}. It
has been used in the study of highly boosted $W$ and Higgs jets
\cite{Wjmass} as well as for boosted tops \cite{Brooijmans}.  In 
terms of the splitting $M \rightarrow A B$, $d_{\rm cut}$ is
equivalent to   
\be
d_{\rm cut} = \min(p_{TA}^2, p_{TB}^2) \Delta R^2_{AB}, \qquad \Delta
R_{AB}^2 \equiv (\phi_A - \phi_B)^2 + (\eta_A - \eta_B)^2, 
\ee
where $p_{Ti}$, $\phi_i$, and $\eta_i$, are the transverse momenta,
azimuth, and pseudo-rapidity of $A$ and $B$ in the lab frame.  The
combination 
\be
z_{\rm cut} \equiv \frac{d_{\rm cut}}{d_{\rm cut} + Q^2_{M}}
\rightarrow  \frac{\min(E_{A}, E_{B})}{E_{A} + E_{B} }
\ee
in the limit that $A$ and $B$ are collinear and massless, so 
$z_{\rm cut}$ is a valid $z$-like variable. 

One nice theoretical property of using $z_{\rm cut}$ as a
discriminating variable compared to $d_{\rm cut}$ directly, is that
$d_{\rm cut}$ measures both the soft ($z \rightarrow 0$) and collinear
($Q^2 \rightarrow 0$) singularities of QCD, whereas $z_{\rm cut}$ only
captures the soft singularity.  Therefore, we expect the $z_{\rm
 cut}$-dependence to pseudo-factorize from the $Q^2$-dependence,
while $d_{\rm cut}$ is expected to be more correlated with $Q^2$.  Of
course, once one isolates a sample of jets whose
invariant mass is near a fixed value $m_t$, $z_{\rm cut}$
and $d_{\rm cut}$ perform nearly identically as discriminating
variables for boosted tops.   
Experimentally, the reason to choose $z_{\rm cut}$ over $d_{\rm cut}$
would be if it is indeed more orthogonal to $Q^2$, both in spectrum and in
systematics. 

There are Lorentz-invariant $z$ variables such as
\be
z_{\rm LI} = \frac{\min(p_{\rm ref} \cdot p_A,p_{\rm ref} \cdot
  p_B)}{p_{\rm ref}\cdot(p_A + p_B)}, 
\ee
where $p_{\rm ref}$ is any reference four vector.  It is easy to check
that when $p_A$ and $p_B$ are massless and collinear, this yields a
valid $z$-like variable as long as $p_{\rm ref}\cdot(p_A + p_B)$ is
not exactly zero.  An obvious choice for $p_{\rm ref}$ is $p_A + p_B$
itself, but a downside to that choice is then $z_{\rm LI}$ can only go 
to zero when $m_A$ or $m_B$ goes to zero, so for typical fat QCD jets with many
hadrons, $z_{\rm LI}$ never sees the QCD soft singularity.  In the
next section, we will take $p_{\rm ref}$ to be one of the proton beams
$\{7 \TeV, 0,0,7 \TeV\},$\footnote{Of course, the choice of
    the magnitude of $p_{\rm ref}$ does not affect our results.}
though this choice  is not particularly 
well-motivated since it violates the reflection symmetry of the
experiment.  Note that $z_{\rm cell}$ is equivalent to choosing
$p_{\rm ref} = \{14 \TeV, 0,0,0\}$. 

We emphasize that in the context of \eq{eq:simpsplitqcd}, $z_{\rm
  cell}$, $z_{\rm cut}$, $z_{\rm LI}$, and all other $z$-like
variables are completely equivalent.  Because there is no preferred
treatment of the virtualities $Q_A^2$ and $Q_B^2$ at the leading
logarithmic level, we cannot say whether 
$z$ should be a Lorentz-invariant function, or just a function of
energy, or just a function of three-vectors.  We also emphasize that
one cannot decide which is the best definition of $z$ from a showering
Monte Carlo alone, because one would be sensitive to the momentum
reshuffling assumptions of the algorithm. 

\subsection{Study of $z$ Variable}
\label{subsec:study-z}

We now study the $z$ variables discussed in the previous section. Hard
scattering, resonance decay, parton showering, and hadronization are performed with
various versions of \pythia\ \cite{pythia}.\footnote{For simplicity,
  we have not considered underlying event or pile-up, though these
  effects are expected to degrade the performance of any jet-based
  measurement.}  We have used the \fastjetversion\ \cite{fastjet}
package to cluster and analyze the jets.  No smearing of the jet or
lepton energies have been included in this study.  Because we will be
using a clustering scheme to define observables, we have approximated
the effect of detector resolution by including finite calorimeter
segmentation.  Hadrons are binned into $\delta\phi \times \delta \eta
= 0.1 \times 0.1$ segments, and each segment is treated as a massless
four-vector whose energy is determined by the hadronic energy deposit.
The calorimeter extends to $|\eta| < 3.0$, and no charged track
information is used.  We will focus on the difference in the shapes of
various observables, therefore all of the kinematical distributions
from different processes are normalized to 1.  

In all the analysis, we only use information about the hardest jet,
and to find that jet we use the anti-$k_T$ algorithm \cite{anti-kt},
which gives nearly conical jets, with $R_0 
= 1.0$.  As mentioned already, this value of $R_0$ is larger than the
typical choices of 0.7 or 0.5, but is justified for the purposes of
studying fat jets.  In a more sophisticated analysis, 
one would choose different values of $R_0$ depending on the expected
$p_T$ spectrum of 
the boosted top in order to track the degree of collimation of the fat
jet.   In order to define the $z$ variable, we have to identify the
subclusters $A$ and $B$.  Taking only the calorimeter cells that 
formed the hardest jet, we recluster with the exclusive $k_T$
algorithm \cite{kt} to force the formation of exactly 2 jets.  This
defines the full splitting $M \rightarrow AB$ from which we can define
any $z$-like variable.   Note that the jet finding algorithm is not
the same as the jet substructure algorithm.

The source of highly boosted top jets in our study is a heavy
resonance with a open decay mode $X \rightarrow t \bar{t}$.  Because
we only care about the $z$ variable shape within a single jet, the
total cross section and angular correlations for $X$ production is
irrelevant, though for definiteness we use a spin-1 sequential $Z'$
\cite{zprime_rev}, which is a \pythiaeightversion\ default.  The main
effect of 
this choice is to select the PDFs used for production, which affects
the $\eta$ distribution of the boosted tops.  We use three benchmark
$X$ masses, and corresponding cuts on the hardest jet $p_T$: 
\be
\label{eq:resonancemassandcut}
\begin{tabular}{c|c}
Resonance Mass & $p_T$ Cut \\
\hline
2 TeV & \phantom{0}800 GeV \\
3 TeV & 1200 GeV \\
4 TeV & 1600 GeV
\end{tabular}
\ee

For the QCD dijet background, we use three different versions of \pythia\ to estimate typical variations among parton shower algorithms:  
\be
\label{eq:pythiaversions}
\begin{tabular}{l|l}
Name & Description \\
\hline
\pythia\ 8 ($p_T$) & \pythiaeightversion\ with the default
$p_T$-ordered shower \\ 
\pythia\ 6 ($p_T$) & \pythiasixversion\ with the ``new'' $p_T$-ordered
shower (\texttt{PYEVNW})\\ 
\pythia\ 6 ($Q$) &  \pythiasixversion\ with the ``old'' $Q$-ordered
shower (\texttt{PYEVNT}) 
\end{tabular}
\ee
We only consider light quark ($uds$) and gluon dijet production,
though the inclusion of heavy-flavor ($cb$) does not dramatically
change the distributions.  Though SM $t\bar{t}$ production is an
important background to $X \rightarrow t \bar{t}$, we consider
high-$p_T$ tops to be part of the signal for this study, and therefore
do not include its contribution since the ratio between SM $t\bar{t}$
and BSM $t\bar{t}$ production is unknown.  Because we are looking only
at the hardest jet for analysis, the QCD background considered here is
a reasonable proxy for any background with the jet $p_T$ cut from
\eq{eq:resonancemassandcut}. 

\begin{figure}
\begin{center}
\includegraphics[scale=0.6]{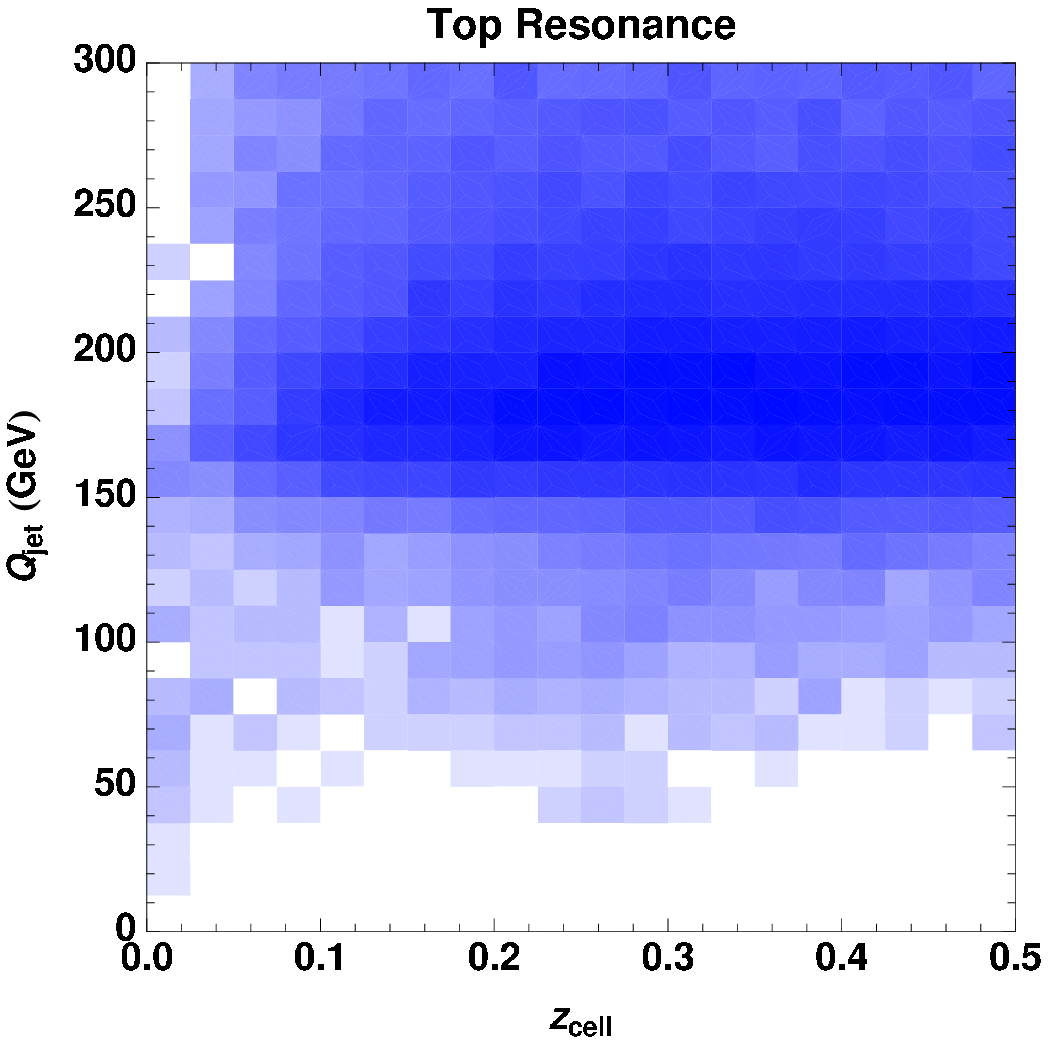} $\qquad$
\includegraphics[scale=0.6]{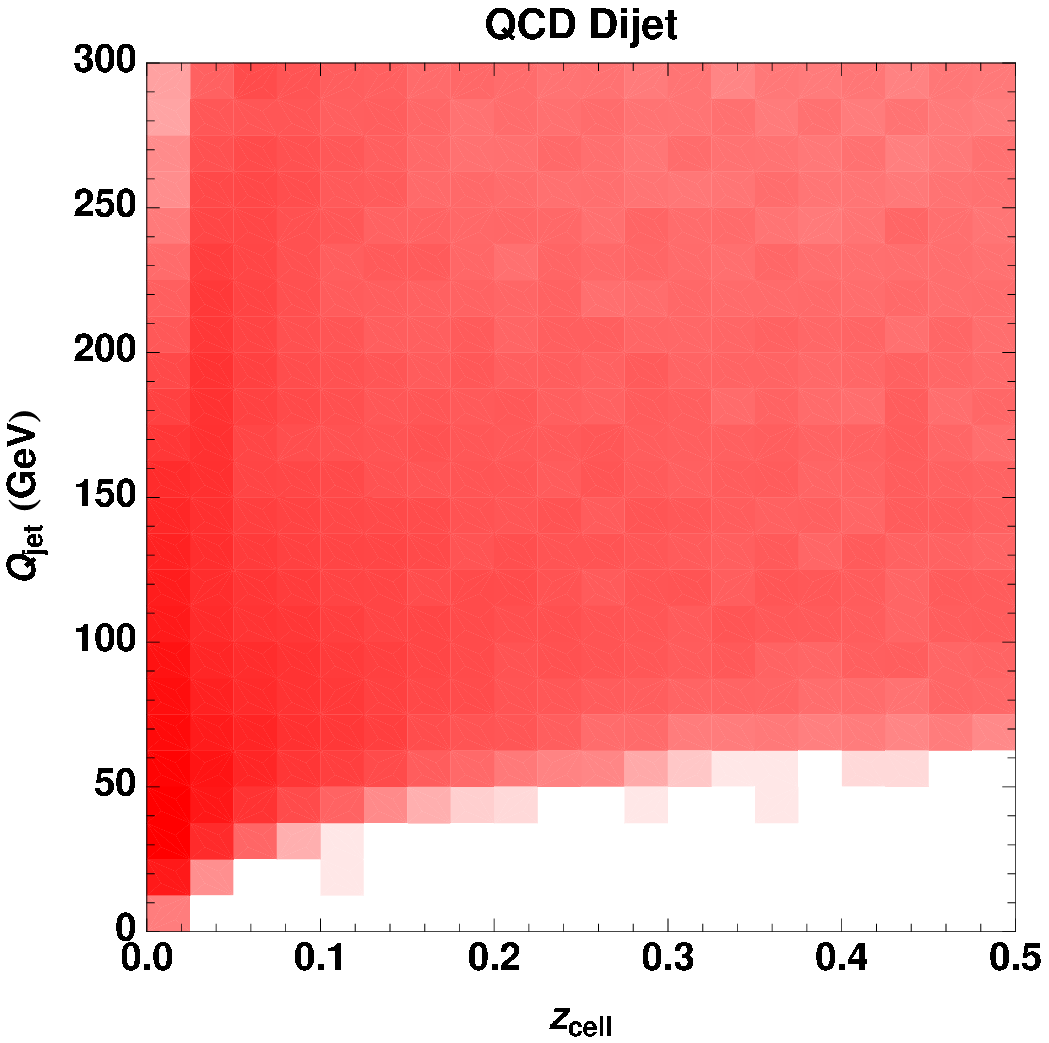} 
\end{center}
\caption{\label{fig:2d-jmass-z} Two dimensional distributions of
  $z_{\rm cell}$ (horizontal axis) versus jet mass $Q_{\rm jet}$
  (vertical axis) for the hardest jet with a 1200 GeV $p_T$ cut.
  Left: boosted top jets from a 3 TeV resonance. The top quark
  resonance structure is visible as a dense horizontal band around
  $Q_{\rm jet} \sim m_t$. Right: QCD dijet production. The infrared
  singularity is visible as $z \rightarrow 0$, though this feature is
  somewhat washed out at large $Q_{\rm jet}$.  The intensity of the
  shading is proportional to the logarithm of the occupancy.} 
\end{figure}

In \fig{fig:2d-jmass-z}, we show two-dimension distributions for
$\{Q_{\rm jet}, z_{\rm cell}\}$ for the 3 TeV top resonance and QCD
dijets, both with a 1200 GeV $p_T$ cut on the hardest jet.  The top
resonance near 170 GeV is visible in the signal sample, and the
collinear and soft singularity near $Q^2,z \rightarrow 0$ is visible
in the QCD sample.  If the clustering procedure were to correctly
identify the  $t \rightarrow bW$ decay, then the expected $z$
distribution from \eq{eq:simpsplitnwa}  would be nearly flat for the
boosted top sample.  Instead, we see that finite detector size and
decay product overlap yields a $z_{\rm cell}$ distribution which falls
off approximately linearly towards $z_{\rm cell} \rightarrow 0$.  In
the QCD sample, the predicted $1/z$ behavior from \eq{eq:simpsplitqcd}
folded with the linear $z_{\rm cell}$ suppression flattens the $z_{\rm
  cell}$ distribution near $Q_{\rm jet} = 170 \GeV$, those the QCD
soft singularity is still observable.

\begin{figure}
\begin{center}
\includegraphics[scale=0.6]{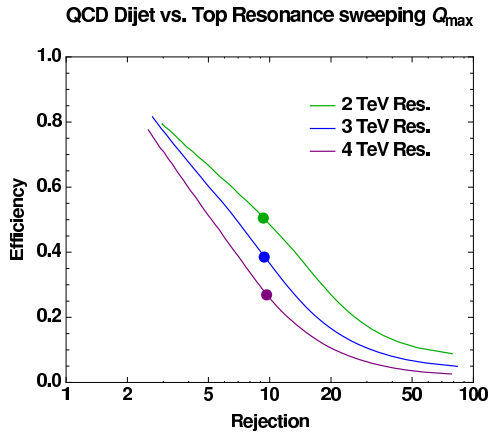}
\end{center}
\caption{\label{fig:EffRejHadQ} Signal efficiency versus background
  rejection using a top mass window selection criteria $160 \GeV <
  Q_{\rm jet} < Q_{\rm max}$ with varying $Q_{\rm max}$.  The
  background is QCD dijets from \pythiaeightversion, and the signals
  and jet $p_T$ cuts are given in \eq{eq:resonancemassandcut}.  In
  this and the remaining efficiency vs.\ rejection plots, rejection
  $r$ is defined as the fraction $1/r$ of the background remaining
  after the cut.  The dots indicate the fiducial value $Q_{\rm max} =
  200 \GeV$.} 
\end{figure}

To isolate a boosted top sample, we first want to impose a jet
invariant mass cut around top mass window $Q_{\rm min}< Q_{\rm jet} <
Q_{\rm max}$.  Because we are using a largish value of $R_0$, the top
decay products are almost always clustered into a single jet, and
choosing a fixed lower bound of $Q_{\rm min} = 160 \GeV$ only
eliminates around 20\% of the signal.  The consequence of using such a
large jet area is that additional hadronic activity from initial state
radiation can enter the jet, so the signal has a long tail to larger
values of $Q_{\rm jet}$.   In \fig{fig:EffRejHadQ}, we vary the value
of $Q_{\rm max}$ and plot the rejection of the QCD background relative
to the signal efficiency.  For comparison with subsequent studies where we include additional variables, we choose a fiducial value of $Q_{\rm max} = 200 \GeV$, labelled by the dots in \fig{fig:EffRejHadQ}.  Amusingly, this value of $Q_{\rm max}$ gives the same 90\% background rejection for all three benchmark $p_T$ cuts, which can be understood from \fig{fig:jmass-pt} because the normalized $Q_{\rm jet}$ distributions intersect near $170 \GeV$.  As the resonance mass increases, the signal efficiency decreases, which is expected because contamination from initial state radiation is a bigger effect at larger parton collision energy, pushing the invariant mass of the top jet out of the top mass window.   We expect that in a realistic
study, one would optimize $Q_{\rm min}$ and $Q_{\rm max}$ in
conjunction with $R_0$.

\begin{figure}[p]
\begin{center}
\includegraphics[scale=0.6]{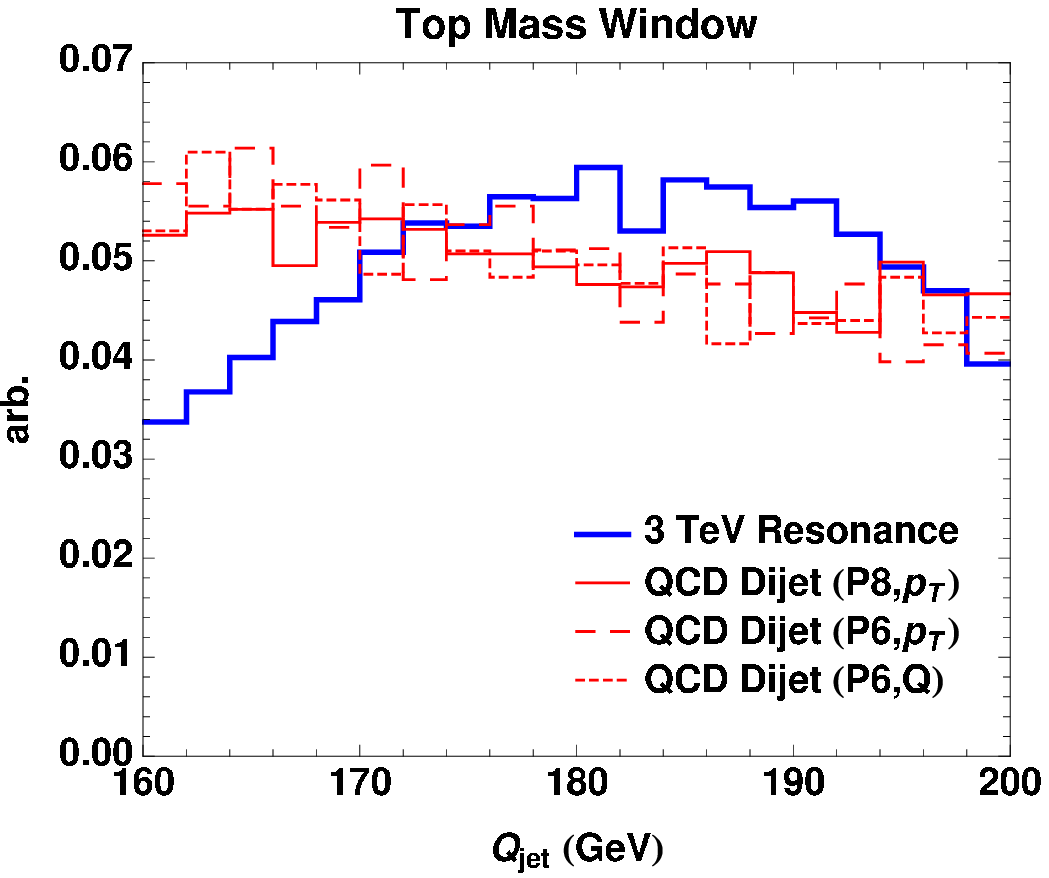} $\qquad$
\includegraphics[scale=0.6]{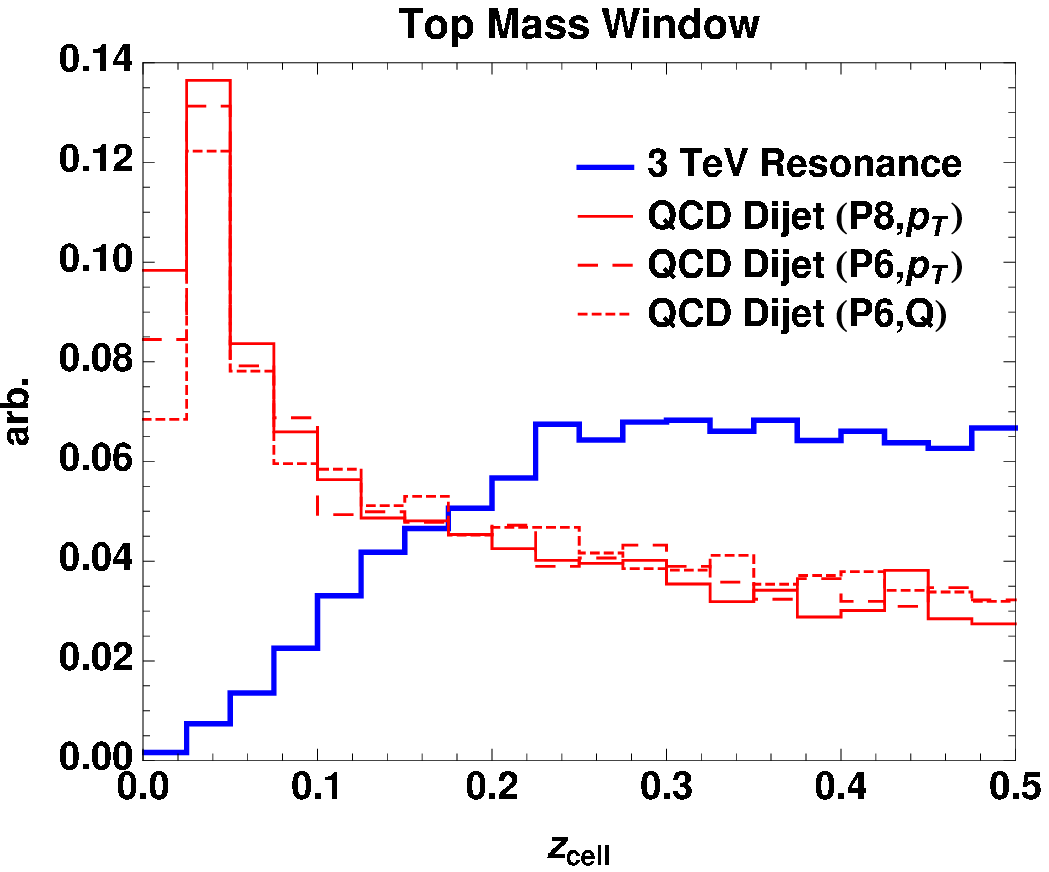}

~\\

\includegraphics[scale=0.6]{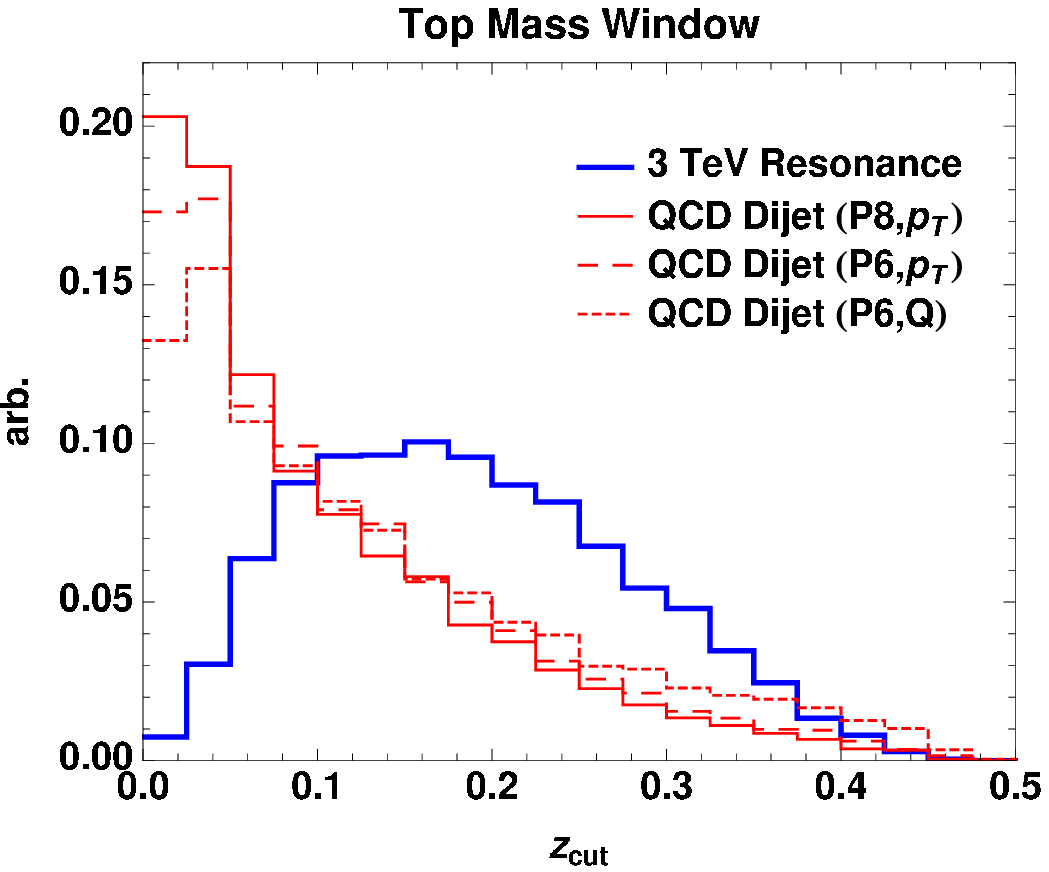} $\qquad$
\includegraphics[scale=0.6]{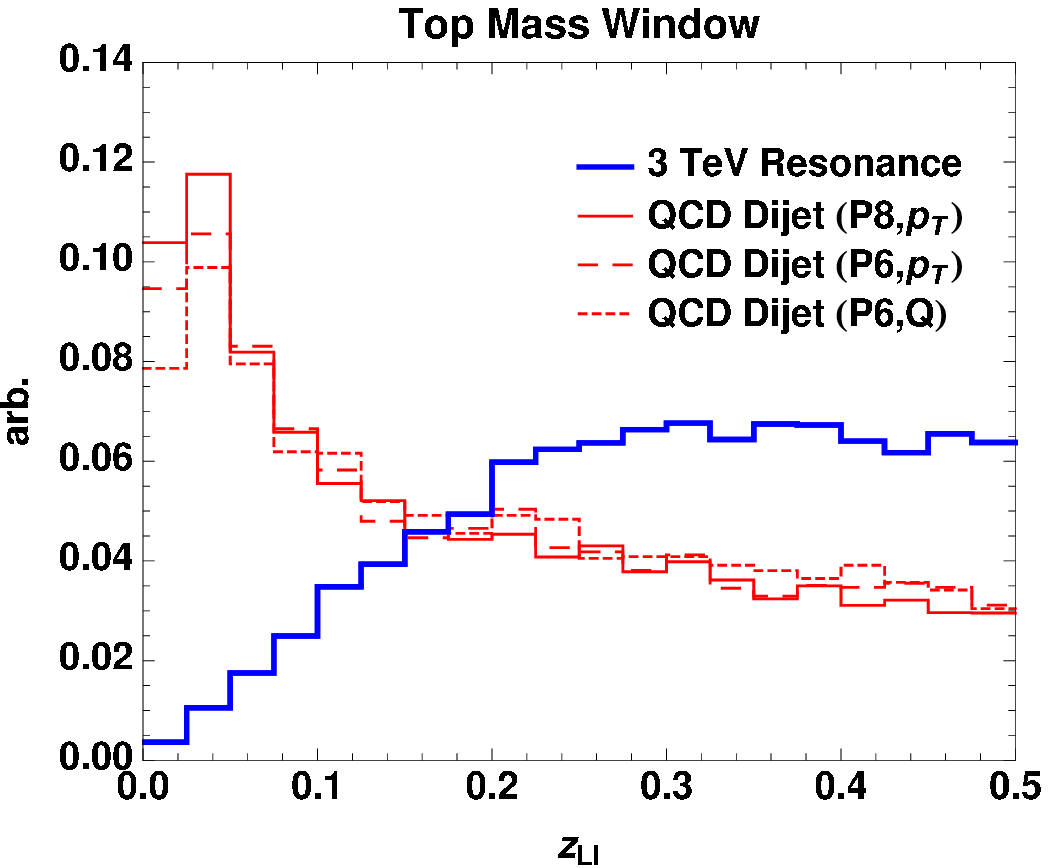} 
\end{center}
\caption{\label{fig:z3TeV} Fat jet spectra of the hardest jet after
  applying a top mass window $160 \GeV < Q_{\rm jet} < 200 \GeV$ cut,
  comparing a 3 TeV top resonance with the QCD dijet background, both
  with a 1200 GeV $p_T$ cut.  From left to right, top to bottom:  jet
  mass $Q_{\rm jet}$, $z_{\rm cell}$, $z_{\rm cut}$, and $z_{\rm LI}$.
  All three \pythia\ versions give similar spectra, validating our
  assumption that $z$ variables should be relatively insensitive to
  showering assumptions.  The dijet background peaks at $z = 0$, while
  the boosted top signal falls off linearly with $z$, so a $z$ cut
  will provide additional discrimination power beyond jet mass alone. 
}
\end{figure}

\begin{figure}[p]
\begin{center}
\includegraphics[scale=0.6]{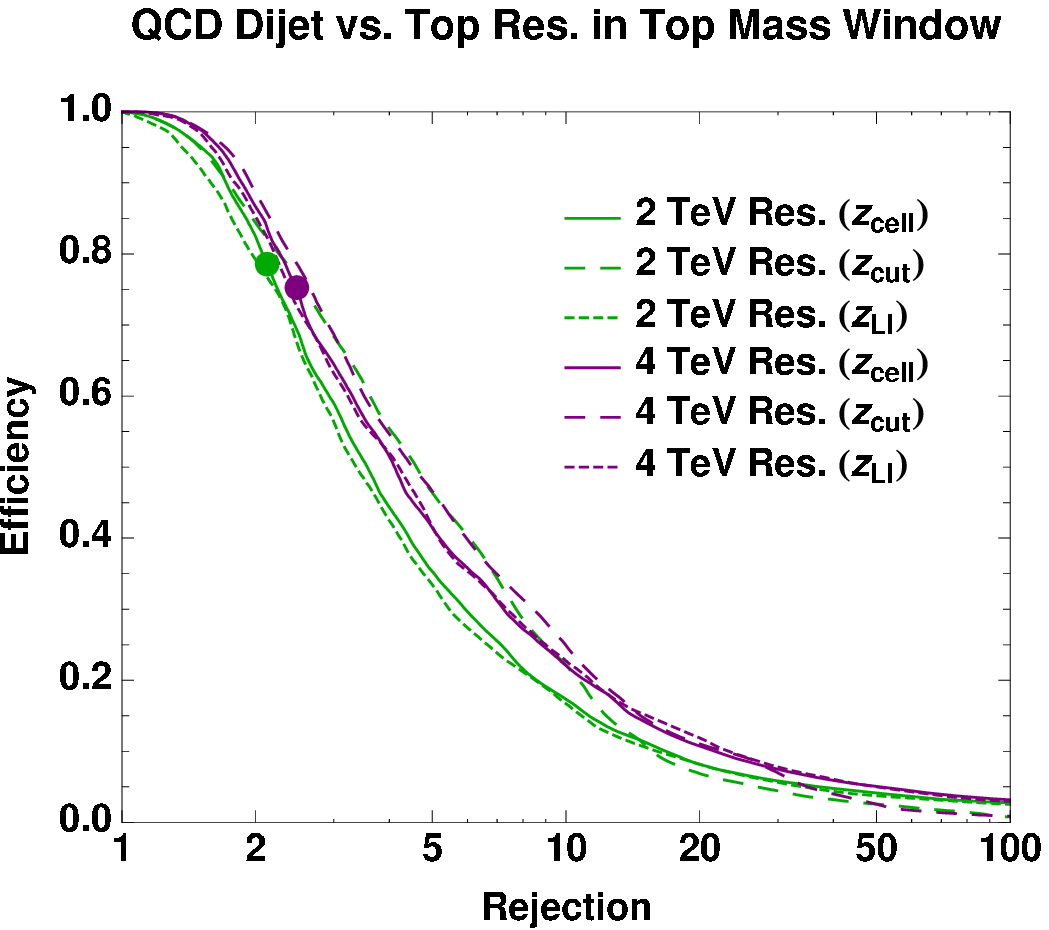} $\qquad$
\includegraphics[scale=0.6]{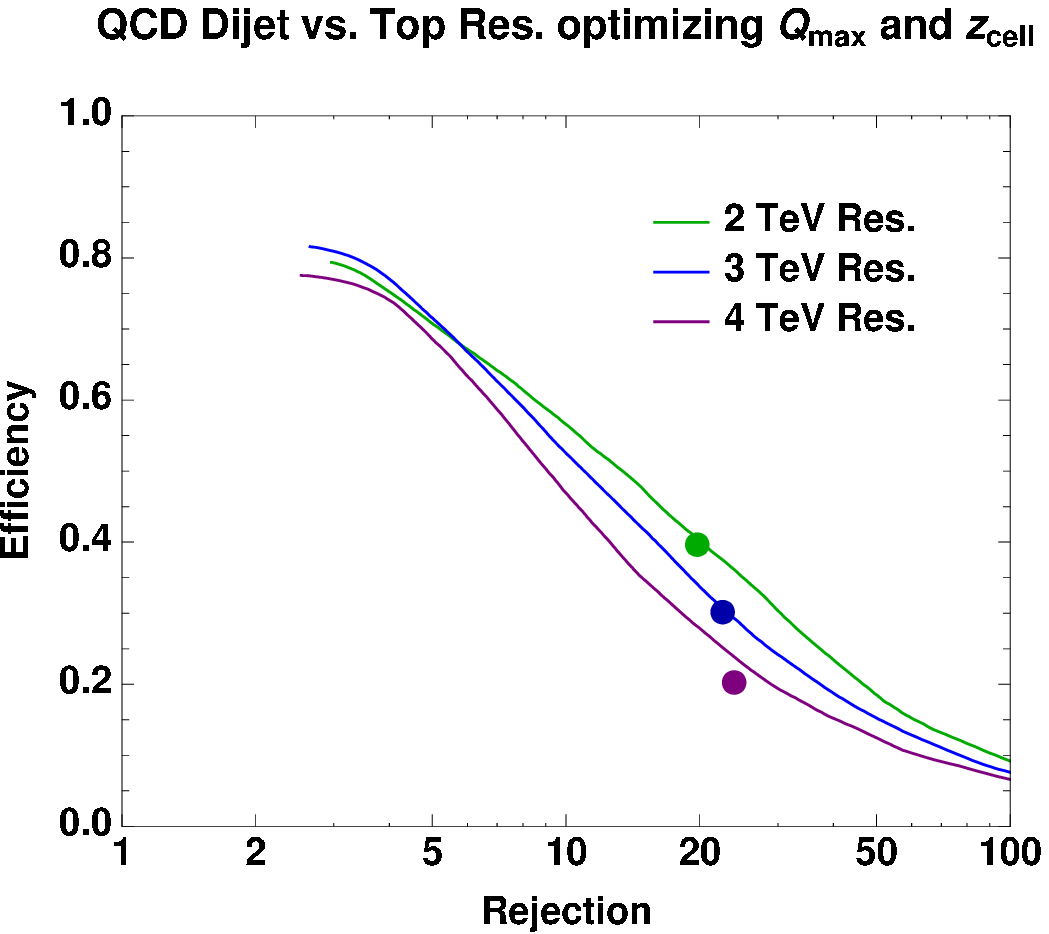}
\end{center}
\caption{\label{fig:eff-rej-z}  Signal efficiency versus background
  rejection using various $z$ variables.  Left:  efficiency
  vs. rejection relative to the $160 \GeV < Q_{\rm jet} < 200 \GeV$
  top window cut.  The dots indicate the fiducial value $\{Q_{\rm
    max},z_{\rm cell}\} = \{200 \GeV, 0.2\}$.  Right:  optimizing the
  $Q_{\rm max}$ and $z_{\rm cell}$ cuts.  This plot is directly
  comparable to \fig{fig:EffRejHadQ}. 
}
\end{figure}

\begin{figure}[p]
\begin{center}
\includegraphics[scale=0.6]{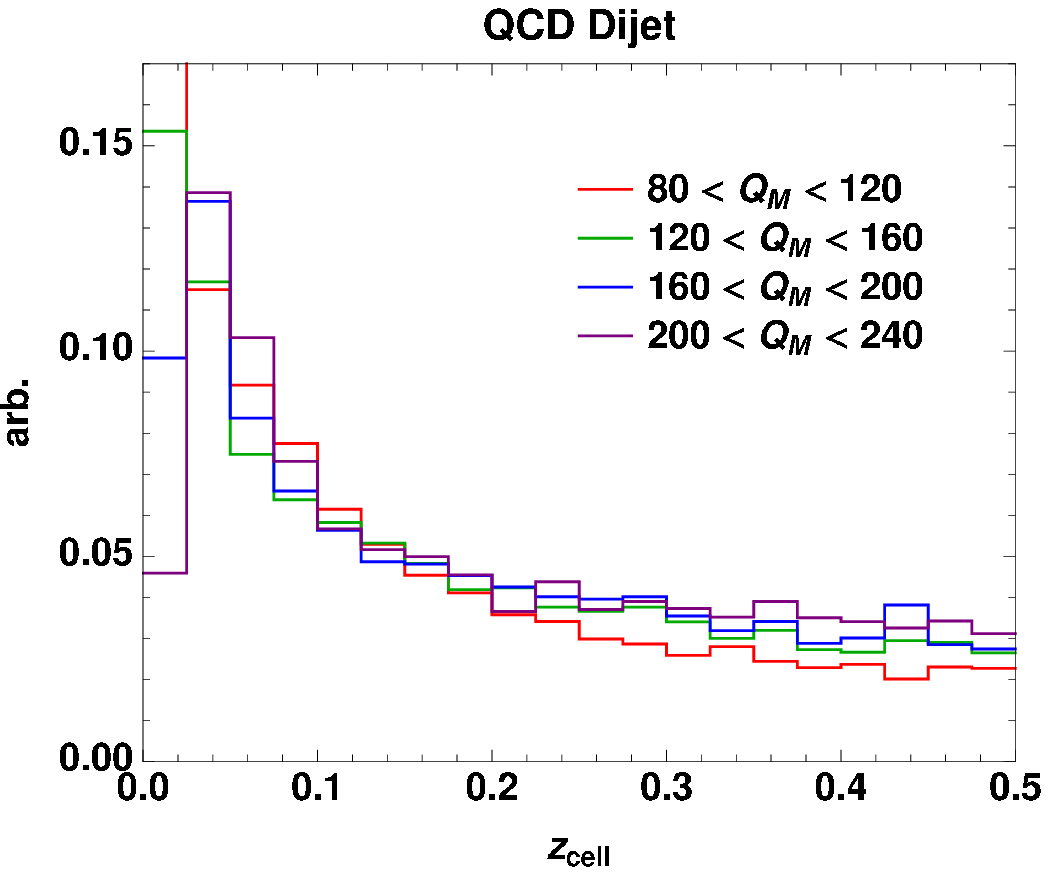} $\qquad$
\includegraphics[scale=0.6]{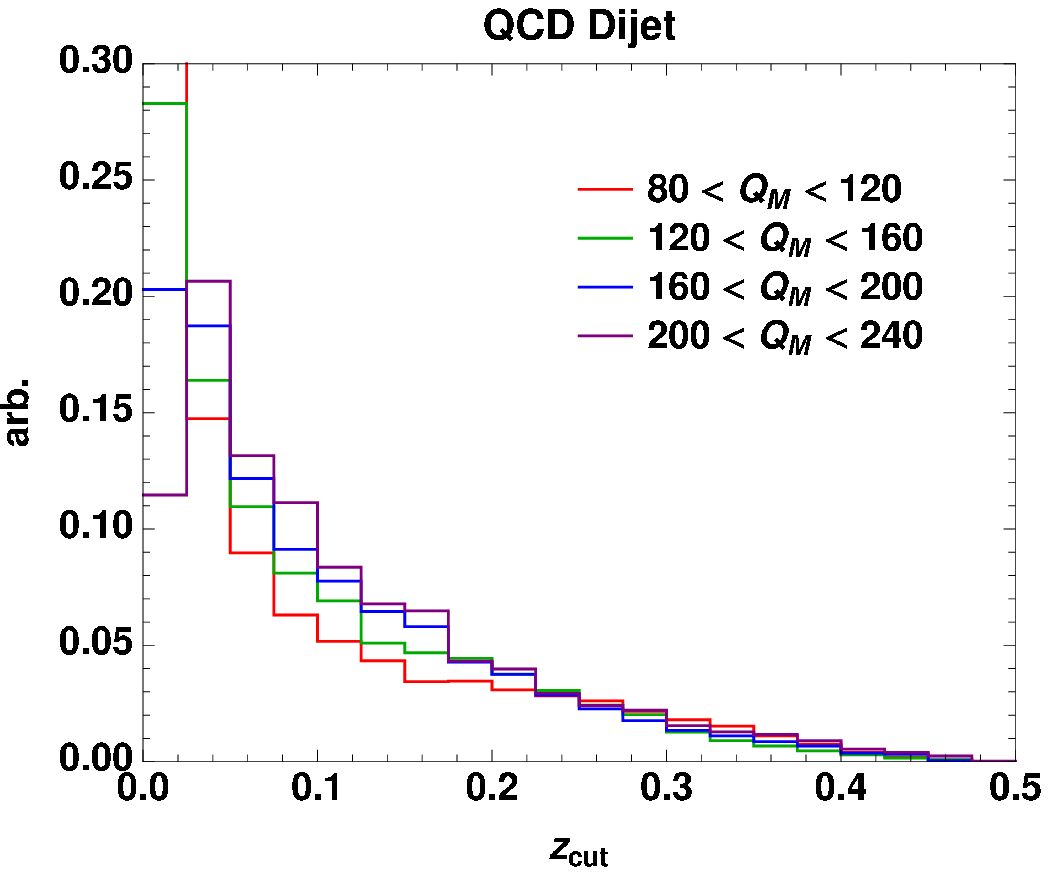} 
\end{center}
\caption{\label{fig:sideband} Evaluating the potential for a sideband
  study of $z$ variables on QCD dijet production.  While the strength
  of the soft singularity of QCD changes with different jet invariant
  mass cuts, the variation is modest, suggesting that the $z$ spectrum
  can be measured at lower $Q_{\rm jet}$ and extrapolated to the top
  mass window.  Left to right:  $z_{\rm cell}$ and $z_{\rm cut}$. 
}
\end{figure}

Having identified the top mass window $160 \GeV < Q_{\rm jet} < 200
\GeV$, we can use the $z$-variables to further purify the sample.  We
demonstrate typical $z$ variable 
distributions in \fig{fig:z3TeV}, again for the 3 TeV
resonance benchmark.  The jet invariant mass is shown in
the upper left panel, and while one might be able to improve signal
to background by tightening the top window criteria, invariant mass
alone probably will not be completely effective in discriminating
boosted top jets from QCD jets.  The $z$-like
variables discussed in the previous section, $z_{\rm cell}$, $z_{\rm
  cut}$ and $z_{\rm LI}$, all show a shape difference between boosted
tops and QCD jets, with the QCD soft singularity clearly active.  In
\fig{fig:eff-rej-z}, we assess the additional discrimination power
offered by the $z$ variable, and for a modest decrease in signal
efficiency, one can increase the background rejection by a factor of 2
to 5 compared to an invariant mass cut alon.
  Instead of fixing the mass window, it is possible to achieve better
  results by optimizing $Q_{\rm max}$ and $z_{\rm cell}$
  together. For example, from the right panel of
  Fig.~\ref{fig:EffRejHadQ}, we see that in the case of 4 TeV
  resonance, the point with fiducial value $Q_{\rm max}=200$ GeV is
  somewhat below the curve from combined optimization. This is fairly
  common in other cases where we combine several variables.

The estimate of the QCD background shape depends on the modeling of
the parton shower process, so it is encouraging that in
\fig{fig:z3TeV},  the three different showering algorithms give
roughly the same $z$ distributions.  This suggests that the structure
of the soft singularity is relatively insensitive to the showering
assumptions, though this insensitivity is somewhat artificial since
all three versions of \pythia\ use similar momentum reshuffling
procedures, and therefore have the same preferred notion of $z$.  In
\pythia\ 6 (Q), one can change the interpretation of $z$ in the shower
using the \texttt{MSTP(43)} option, but we found that these variations
were no larger than the variations between different
\pythia\ versions.

A realistic evaluation and optimization of the $z$-variable should be
performed once we can study real QCD jets at the LHC.  It is
important to be able to perform such a study in a ``side-band'', i.e.\ 
in a kinematical regime where the presence of top jets does not
distort the infrared structure of QCD jets.\footnote{Away from $Q_{\rm
    jet} \sim m_t$, there could be contamination from the signal if
  the top was far off-shell or if the supposed ``top jet'' only
  contained part of the top quark decay products.  Because $z$
  distribution for a boosted top will be determined by a hard prompt
  decay, we expect the signal to be less sensitive to the theoretical
  uncertainties associated with parton shower modeling. Therefore any
  signal contamination effects could be estimated from Monte Carlo.} 
 In Fig.~\ref{fig:sideband}, we show the $z$-variable distribution for
 QCD jets for different 
jet mass windows.  The variation in the $z$-distributions are mild across
a wide range of jet masses, with the biggest difference being the peak
shape near $z \rightarrow 0$, suggesting that a side-band study should
be feasible. 

\section{Multi-Body Prospects}
\label{sec:other}

\subsection{Theoretical Challenges}

In the previous section, we looked at variables that approximated jet
fragmentation as a simple two-body decay.  Because we used a parton
shower Monte Carlo to generate events, and because the parton shower
is defined by a sequence of $1 \rightarrow 2$ splittings, we argued
that the variables most likely to be correctly described by the Monte
Carlo are ones that only involve a single $1 \rightarrow 2$ split.
Given the momentum reshuffling ambiguity as to how to treat successive
emissions in the parton shower, we wanted our analysis to be as
independent of subsequent splittings as possible. 

Because an $M \rightarrow AB$ splitting is characterized completely by
$Q_M^2$ and $z_M$, many seemingly ``better'' observables are really
equivalent to the analysis of \Sec{sec:tophadronic} with a different
choice of $z$ variable.  We already saw that there was a $z_{\rm cut}$
variable that captured the physics of a $d_{\rm cut}$ measurement.
Similarly, an energy ``dipole'' variable is just a different way of
phrasing $\{Q_M^2,z_M\}$.  In principle, one could gain leverage by
using information about the daughter virtualities, but already the
different choices for $z$ differ precisely by the treatment of $Q_A^2$
and $Q_B^2$, so some of the daughter information is already contained
in $z$.   The only way to dramatically improve an analysis based on $M
\rightarrow AB$ kinematics is to find experimental observables that do
a better job correctly identifying the relevant subclusters $A$ and
$B$. 

Given the limitations of two-body observables, one would like to look
at multi-body signatures.  Indeed, the decay of a top quark looks less
like a $1 \rightarrow 2$ splitting and more like a $1 \rightarrow 3$
splitting, so one should be able to use multi-body information to help
distinguish boosted tops from QCD jets.  Such variables are not very
well understood in the parton shower approximation, but in principle
one could extend \eq{eq:simpsplitqcd} to correctly handle multiple QCD
emissions.  Training such variables at the LHC with QCD
  jets in some proper control region will certainly provide useful
  information.  With multi-body
information about QCD, one can consider 
more elaborate jet substructure observables, and we will look at two
specific kinds in this section.  

\subsection{Boost-Invariant Event Shape}

While boosted tops might be described theoretically by an $M
\rightarrow ABC$ splitting, one still has to find an experimental
proxy for the $A$, $B$, and $C$ subclusters.  Instead of using a
clustering algorithm, an alternative strategy is to construct an event
shape variable that uses all of the hadrons in a jet to form an
observable that measures the gross energy distribution. 

The goal is to build an event shape that probes the fact the top decay
products are widely separated in the top rest frame, so one wants a boost-invariant event shape.  Ideally, the event shape would
be invariant under both the boost axis and the boost magnitude.
Unfortunately, building a meaningful event shape that is invariant
under choice of boost axis is difficult, because in the $M$ rest
frame, the splitting $M 
\rightarrow ABC$ defines a plane.  If the boost axis is perpendicular
to this plane then $A$, $B$, and $C$ look well-separated, but if the
boost axis is parallel to the plane, then $A$, $B$, and $C$ overlap.   

\begin{figure}
\begin{center}
\includegraphics[scale=0.6]{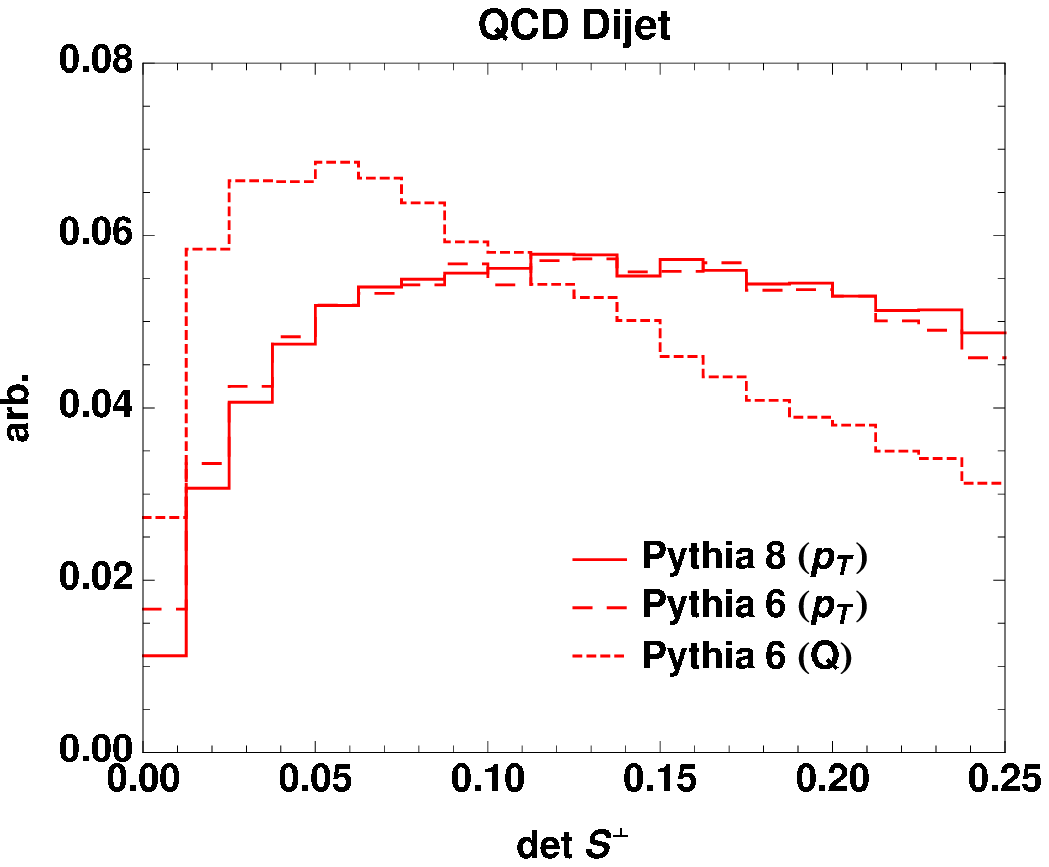} $\qquad$
\includegraphics[scale=0.6]{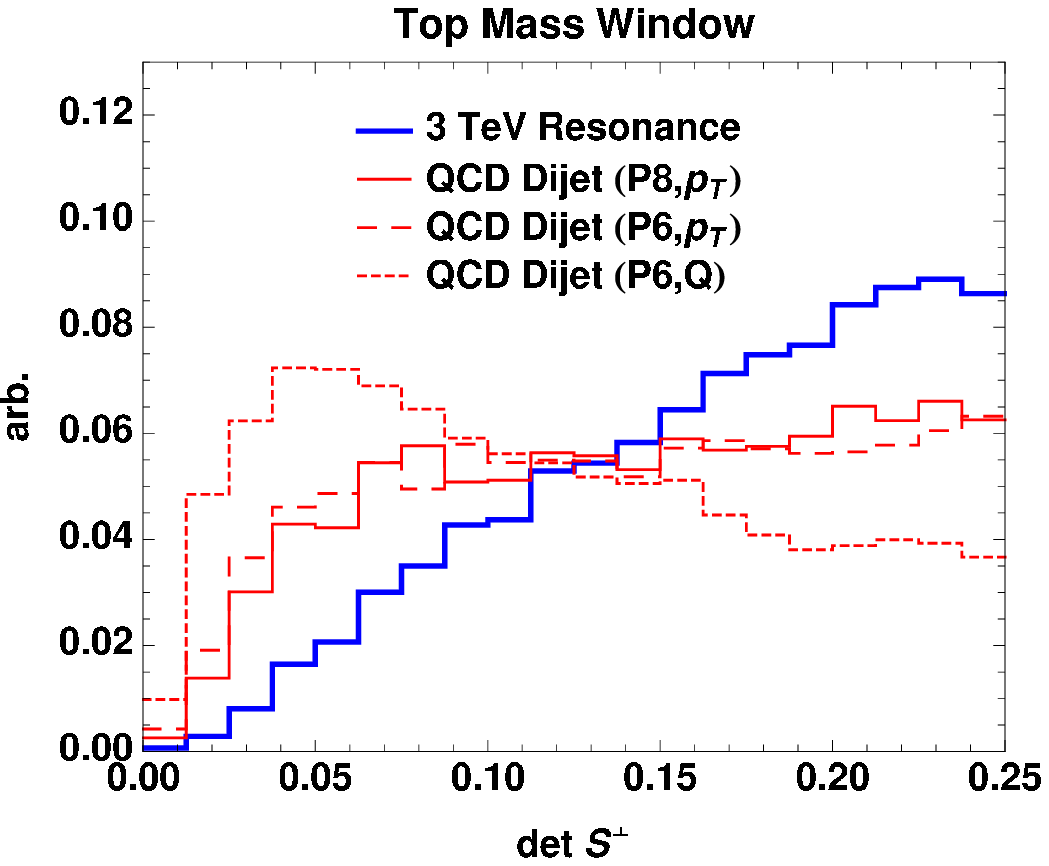}
\end{center}
\caption{\label{fig:detSij} Left:  QCD dijet predictions for $\det
  S^{\perp}$ with a $p_T$ cut of 1200 GeV, showing large variations
  between different shower evolution variables.  Right:  $\det
  S^{\perp}$ after imposing the top window cut $160 \GeV < Q_{\rm jet}
  < 200 \GeV$, comparing to a 3 TeV top resonance.  While $\det
  S^{\perp}$ shows promise in separating boosted tops from QCD fat
  jets, it is difficult to make a firm conclusion given the large
  theoretical variance. 
}
\end{figure}

We can still form an event shape that is invariant under the boost
magnitude, by considering a variant to the ordinary sphericity tensor
\cite{sphericity}.\footnote{Strictly speaking, even this event shape is not
  invariant under boosts given finite calorimetry.  Even though
  $\vec{p}^\perp$ is invariant under boosts, the calorimetry is
  defined by $\phi$ and $\eta$, which is invariant only under boosts
  along the beam axis and not to boosts along the top momentum axis.}
Taking the $z$-axis to be the boost direction, consider a jet with
total four vector $\{E_{\rm 
 jet}, \vec{0}^\perp,p^z_{\rm jet}\}$ and constituents $p^\mu_\alpha
=\{E_\alpha, \vec{p}^\perp_\alpha, p^z_\alpha\}$.   The (linear) jet
transverse sphericity tensor $S^{\perp ij}$ is an object that is
invariant under boosts along the $z$-axis:   
\be
\label{eq:sperp}
S^{\perp ij} = \frac{\displaystyle \sum_{\alpha \in \mathrm{jet}}
 \frac{\vec{p}^{\perp i}_\alpha \vec{p}^{\perp
     j}_\alpha}{|\vec{p}^\perp_\alpha|}}{\displaystyle \sum_{\alpha
   \in \mathrm{jet}}|\vec{p}^\perp_\alpha|}.
\ee
There is only one non-trivial eigenvalue of $S^{\perp}$ since the two
eigenvalues sum to 1, so we will take the determinant of $S^{\perp}$
to be our boost-invariant event shape.  Note that $\det S^{\perp}$ is
identically 0 in the two-body limit, so $\det S^{\perp}$ does not have
a $z$-like interpretation and is a true multi-body distribution.  It
is easy to check that $0 \le \det S^{\perp} \le 0.25$.

One can think of $\det S^{\perp}$ as measuring the degree of
transverse phase space democracy within a jet.  Boosted tops are
peaked at large values of $\det S^{\perp}$ because when the boost axis
is perpendicular to the $ABC$ plane, the three decay products fill the
transverse plane.  QCD jets peak at smaller values of $\det S^{\perp}$
because filling the transverse plane would require two large kicks
relative to the jet axis, but this is suppressed by the splitting
functions. 

While $\det S^{\perp}$ is a nice event shape in principle, it is not
surprising that the theoretical distribution is not well understood in
the parton shower picture.  In the left panel of \fig{fig:detSij} we
show the $\det S^{\perp}$ distribution for the three different
versions of \pythia\ from \eq{eq:pythiaversions}, and one can see that
depending on the choice of evolution variable in the shower, the
prediction for the $\det S^{\perp}$ distribution can vary by 30\%.  In
the right panel of \fig{fig:detSij}, we isolate jets within the top
mass window $160 \GeV < Q_{\rm jet} < 200 \GeV$ and compare to a 3 TeV
top resonance as in \Sec{subsec:study-z}.   Without information about
which prediction is closer to true QCD, it is impossible to say how
effective $\det S^{\perp}$ would be to separate signal from
background. 

\subsection{$W$ Reconstruction}
\label{subsec:wreco}

The challenge to using event shapes is that they tend to probe the
generic properties of jets and therefore require good theoretical
control over QCD.  A more promising approach is to look at observables
that are very distinctive for boosted tops, such that one can be
confident in their effectiveness even without accurate QCD
predictions. 

\begin{figure}[p]
\begin{center}
\includegraphics[scale=0.6]{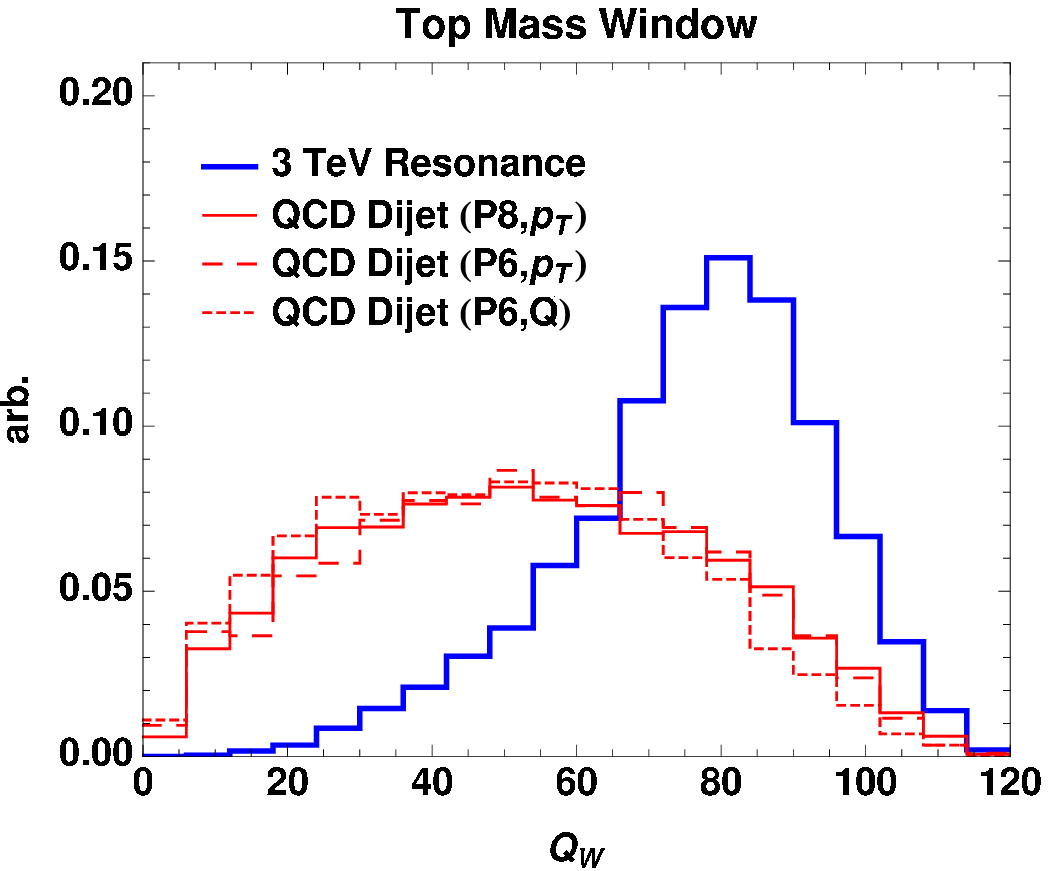}$\qquad$
\includegraphics[scale=0.6]{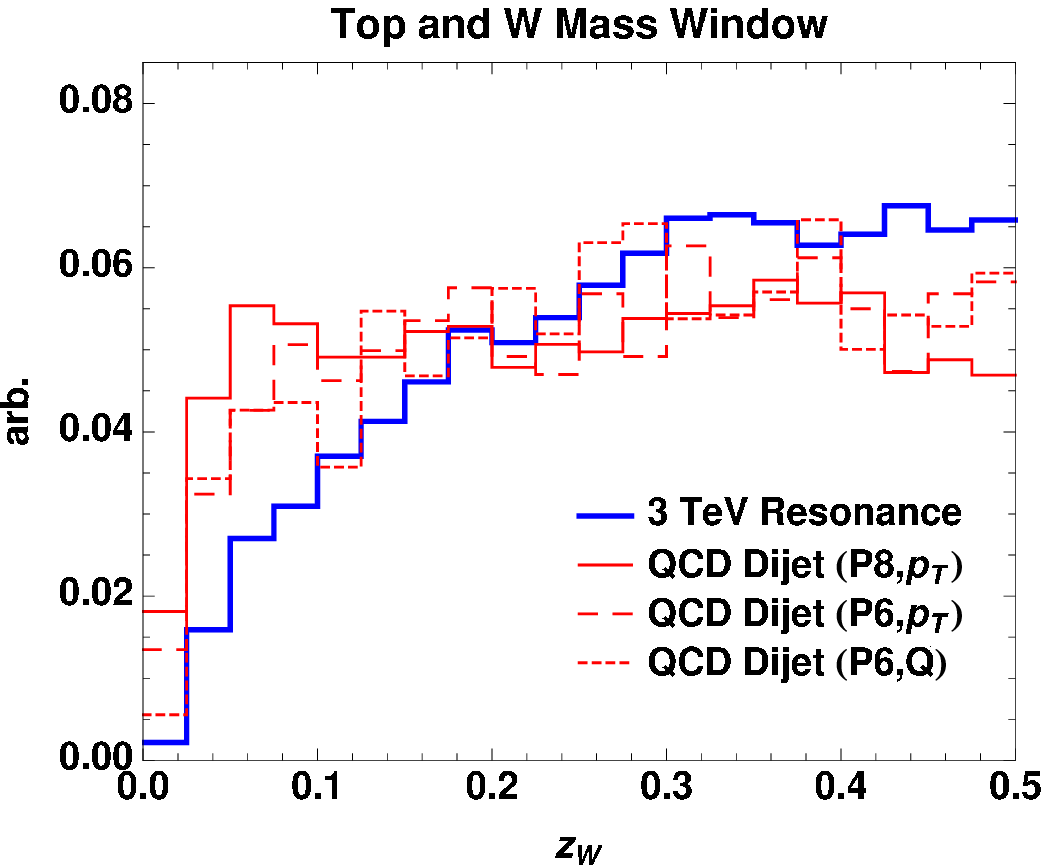}
\end{center}
\caption{\label{fig:wreco} Left: applying the $W$ reconstruction
  technique in the top mass window $160 \GeV < Q_{\rm jet} < 200
  \GeV$.   A clear $W$ resonance in $Q_W$ is seen centered at $80
  \GeV$ for the top resonance, while QCD dijets are peaked at a lower
  value.    Right:  applying an additional $W$ window cut $60 \GeV <
  Q_W < 100 \GeV$ and looking at $z_W$, the analog of $z_{\rm cell}$
  within the candidate $W$.  While there is some evidence of the QCD
  soft singularity as $z_W \rightarrow 0$, it is not strong enough to
  justify applying additional cuts on this variable.} 
\end{figure}

For boosted tops, the obvious distinguishing feature is the presence
of an \emph{on-shell} $W$ inside a jet.  Using a jet clustering
algorithm to identify an $M \rightarrow ABC$ splitting, one can check
whether a pair of $A$, $B$, and $C$ reconstructs an $80 \GeV$
resonance.  In the left panel of \fig{fig:wreco}, we use the same
$k_T$ clustering 
procedure as \Sec{subsec:study-z} to identify the $M \rightarrow ABC$
splitting, and plot the minimum pair-wise invariant mass $Q_W$ between $A$,
$B$, and $C$.  One can see a clear peak at $80 \GeV$ in the signal,
and there is less variation in the different background estimates,
though one should regard that observation with caution.  

\begin{figure}[p]
\begin{center}
\includegraphics[scale=0.6]{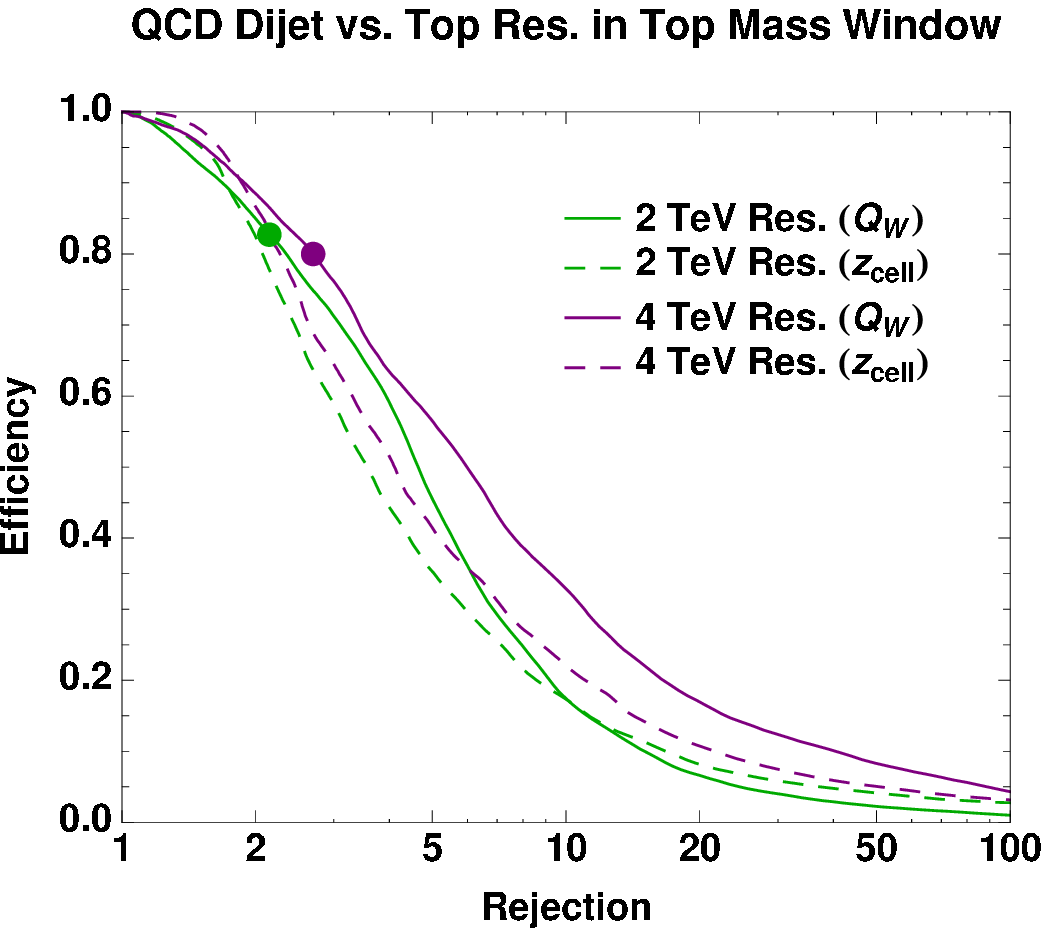} $\qquad$
\includegraphics[scale=0.6]{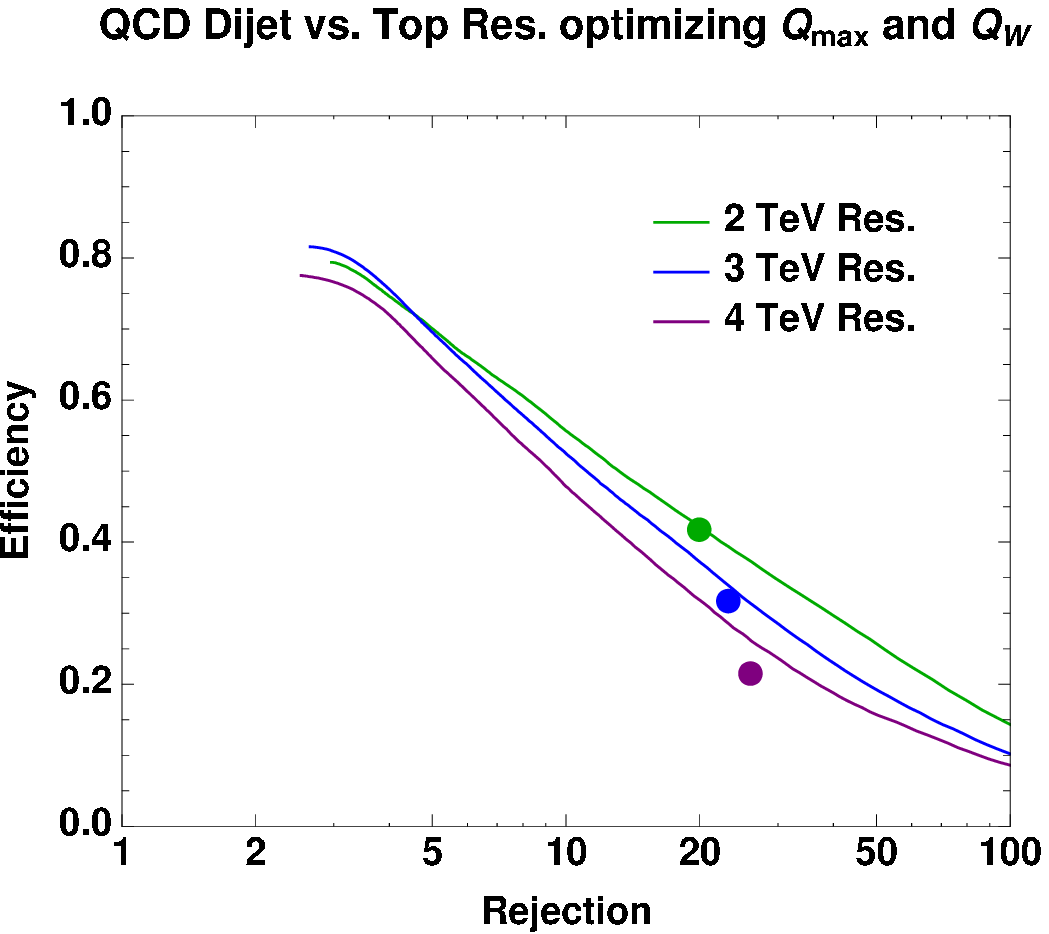}
\end{center}
\caption{\label{fig:wrecoeff}Signal efficiency versus background
  rejection using $W$ reconstruction.  Left:  efficiency vs. rejection
  relative to the $160 \GeV < Q_{\rm jet} < 200 \GeV$ top window cut.
  The $z_{\rm cell}$ distribution is shown for comparison, and the
  dots indicate the fiducial value $\{Q_{\rm max},Q_{W}\} = \{200
  \GeV, 60 \GeV\}$.  Right:  optimizing the $Q_{\rm max}$ and $Q_{W}$
  cuts.  This plot is directly comparable to
  \figs{fig:EffRejHadQ}{fig:eff-rej-z}.} 
\end{figure} 

Once a candidate $W$ is identified, one could try to use the same logic as
\eqs{eq:simpsplitnwa}{eq:simpsplitqcd} to test for the QCD soft
singularity with the supposed $W$ jet.  In the right panel of
\fig{fig:wreco}, we plot the $z_W$ (defined analogously to $z_{\rm
  cell}$) distribution for events with $60 \GeV < Q_W < 100 \GeV$.
While there is a slight difference in shape, it is not nearly as
dramatic as the shape variations seen in \fig{fig:z3TeV}, probably
because the overlap between subclusters is larger, blurring the QCD
soft singularity. 

Therefore, we conclude that the best use of hadronic $W$
reconstruction is through a cut on $Q_W$.  As seen in
\fig{fig:wrecoeff}, $Q_W$ seems to perform as well if not better than
$z_{\rm cell}$ as a probe of boosted tops.  Unfortunately, we have
checked separately that  $Q_W$ and
$z_{\rm cell}$ have correlated rejection rates, so one does not gain
much by combining both cuts, though the systematics might be different
enough to justify using both.  In order to fully assess the
effectiveness of $W$ reconstruction against background, one should
also study the background from boosted hadronic $W$s  \cite{Wjmass}
with additional 
radiation, which will pass the $Q_W$ cut more readily but fail the top
window and $z_{\rm cell}$ criteria more often. 

\section{Leptonic Top Decays}
\label{sec:topleptonic}

\subsection{Stuck Lepton Strategy}

When a top decays leptonically as $t \rightarrow b \ell \nu$, the
neutrino carries away a substantial fraction of the top quark energy.  It is
straightforward to check that in this decay,
the range achievable for $Q^2_{\rm visible} = (p_\ell + 
p_{b})^2$ is 
\be
\label{eq:Qvis}
\frac{m_t^2 m_b^2}{m_t^2 - m_W^2} + \mathcal{O}(m_b^4) < Q^2_{\rm
  visible}  < m_t^2 - m_W^2  + \mathcal{O}(m_b^2), 
\ee
where we have taken the lepton to be massless for simplicity, and
$m_b$ should be regarded as the energy of the $b$ jet including QCD
radiation.\footnote{In Eq.~\ref{eq:Qvis}, we have not
    explicitly included possible missing energy in semileptonic
    $B$ decays.  This is expected to a subleading effect in evaluating $Q^2_{\rm
  visible}$.  We do, however, take this effect into account in our numerical
    simulation.}  Because this range extends down to very low values of 
$Q^2_{\rm visible}$, looking for jets with $Q^2 \sim m_t^2$ is no longer an
efficient strategy for boosted top identification. Still,
  we will use $Q^2_{\rm visible}$ as one selection criteria for
  boosted leptonic tops.  We could 
in principle reconstruct the neutrino momentum and therefore the full
kinematics in the semileptonic channel of QCD $t \bar{t}$ production,
particularly near the threshold.  However, depending on 
the particular top-rich BSM signal one is 
looking for, such reconstruction may not be feasible as there may be additional
sources of missing energy in the event.    
 
Even if one cannot use the invariant mass of visible objects to
completely isolate boosted leptonic tops, the lepton from a boosted
top is not isolated from the top jet hadronic activity, 
and the presence of a hard lepton ``stuck'' in a jet is still quite
unique.  One irreducible background comes from high-$p_T$ bottom and
charm jets as the decay of the resulting heavy flavor meson can
contain leptons.  Another irreducible background is $W/Z +
\mbox{jets}$ where the gauge boson 
decays leptonically and the lepton ``accidentally'' gets caught in a
jet.   There is also a ``background'' to boosted leptonic tops from
boosted hadronic tops with leptonic $b$ decays, though the visible invariant
mass can help separate these signals.   

Because soft-muon tags are already used for $b$-tagging
\cite{atlasTDR,cmsTDR}, we
will refer to the stuck lepton as a muon, though in principle, stuck
electrons might also be observable.\footnote{It is more challenging to
  identify an electron within a jet, although its different energy
  deposit pattern on the electromagnetic calorimeter combining with
  tracking information could be used \cite{chris}.}  While our expectation
is that muons should be relatively clean, potential fakes must be
taken into account.  Two commons ways muon fakes can happen is kaon or
pion decay in flight or if hadronic activity punches-through to the
muon chamber \cite{punch-through}.  In this work, we will not discuss these
reducible backgrounds from lepton misidentification, though this might
be more dangerous than the irreducible backgrounds because of the
large rate for QCD jets.  On the other hand, if the fake muon only
carries a small fraction of the total hadronic energy, then the muon
spectrum will be similar to heavy flavor jets, and we will see that we
have good control over the heavy flavor background.  For the numerical
results presented in the rest of this section, we have not taken into
account realistic lepton efficiencies and fake rates, and leave this
to future study. 

Unlike the case of hadronic tops, where there were large theoretical
uncertainties in the jet substructure observables, we expect the
leptonic variables to be well-modeled by (tuned) Monte Carlo.   For
the irreducible backgrounds, the source of the leptons are either $W$
bosons or $B$ mesons whose decays are well-understood.   This leaves
two main sources of theoretical uncertainty: the invariant mass
spectrum of bottom and charm jets and the spectrum of the $B$ and $D$
meson inside the jet.  The invariant mass spectrum should be
measurable from data, using a sample of $b$-tagged but
soft-muon-vetoed jets.  One should also have some understanding of the
$B$/$D$ meson spectrum from the displaced vertex mass using the same
sample. 

\subsection{Lepton Variable Study}

To identify a candidate leptonic boosted top, we identify the muon
with the hardest $p_T$ in an event, and use the anti-$k_T$ algorithm
with $R_0 = 1.0$ to figure out which hadronic activity is associated
with that muon.  We call the muon four-vector $p_\mu$ and the
remaining hadronic activity $p_b$, though we emphasize that we do not
use $b$-tagging in this study and  the hadronic activity need not
initiated by heavy flavor such as the case of the $W/Z + \mbox{jets}$
background.  We define three leptonic spectrum variables $z_\mu$,
$\Delta R$, and $x_\mu$ and test their effectiveness for background
rejection: 
\be
z_\mu = \frac{E_\mu}{E_\mu + E_b},  \qquad \Delta R = \Delta
R(p_\mu,p_b), \qquad x_\mu  = \frac{2 p_\mu \cdot p_b}{(p_\mu +
  p_b)^2}.
\ee
We will describe the reason for these three variables in more detail
below.  In order to reject truly isolated muons, we impose a cut of
$z_\mu < 0.9$ on all the samples.  As in \Sec{subsec:study-z}, we are
focused on the shapes of the backgrounds, and normalize all
distributions to 1.

\begin{figure}
\begin{center}
\includegraphics[scale=0.6]{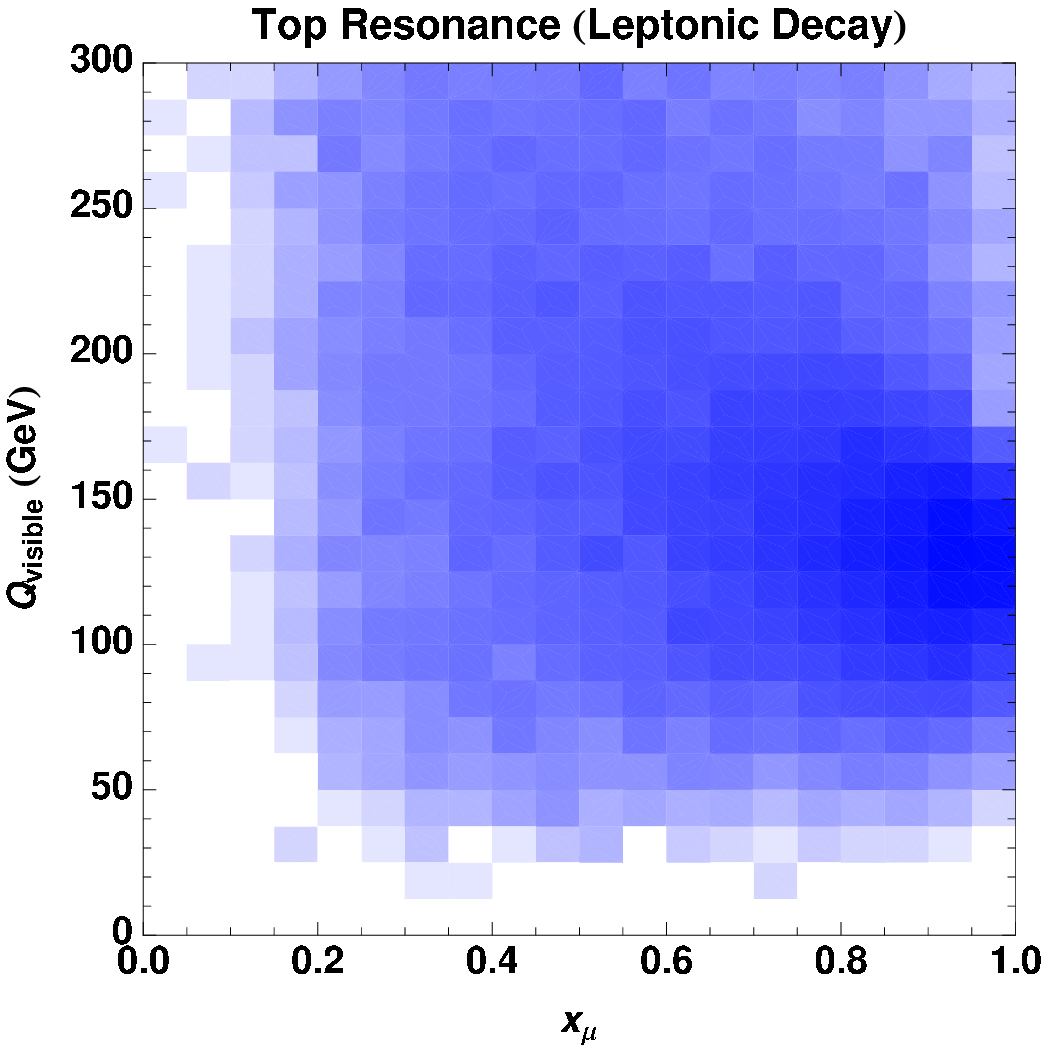} $\qquad$
\includegraphics[scale=0.6]{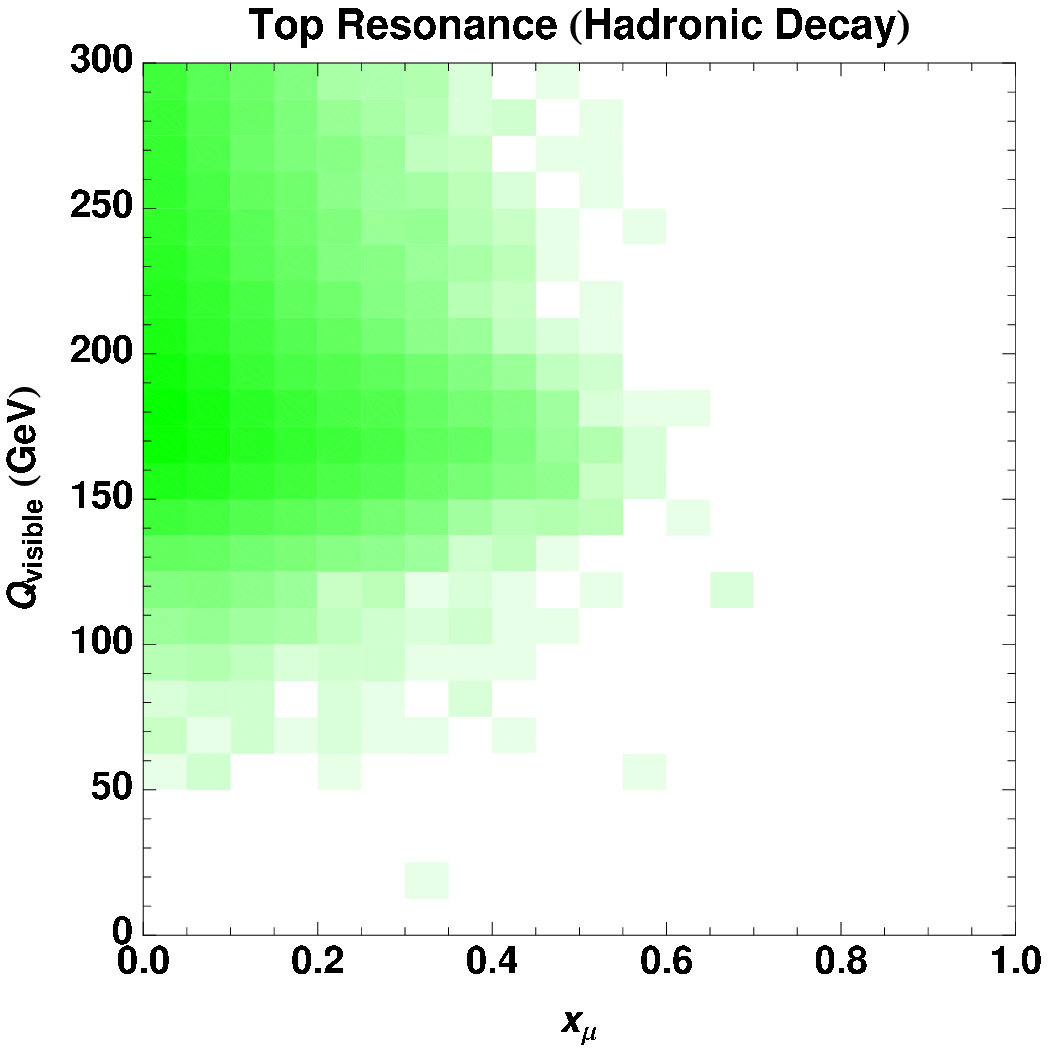}

~\\

\includegraphics[scale=0.6]{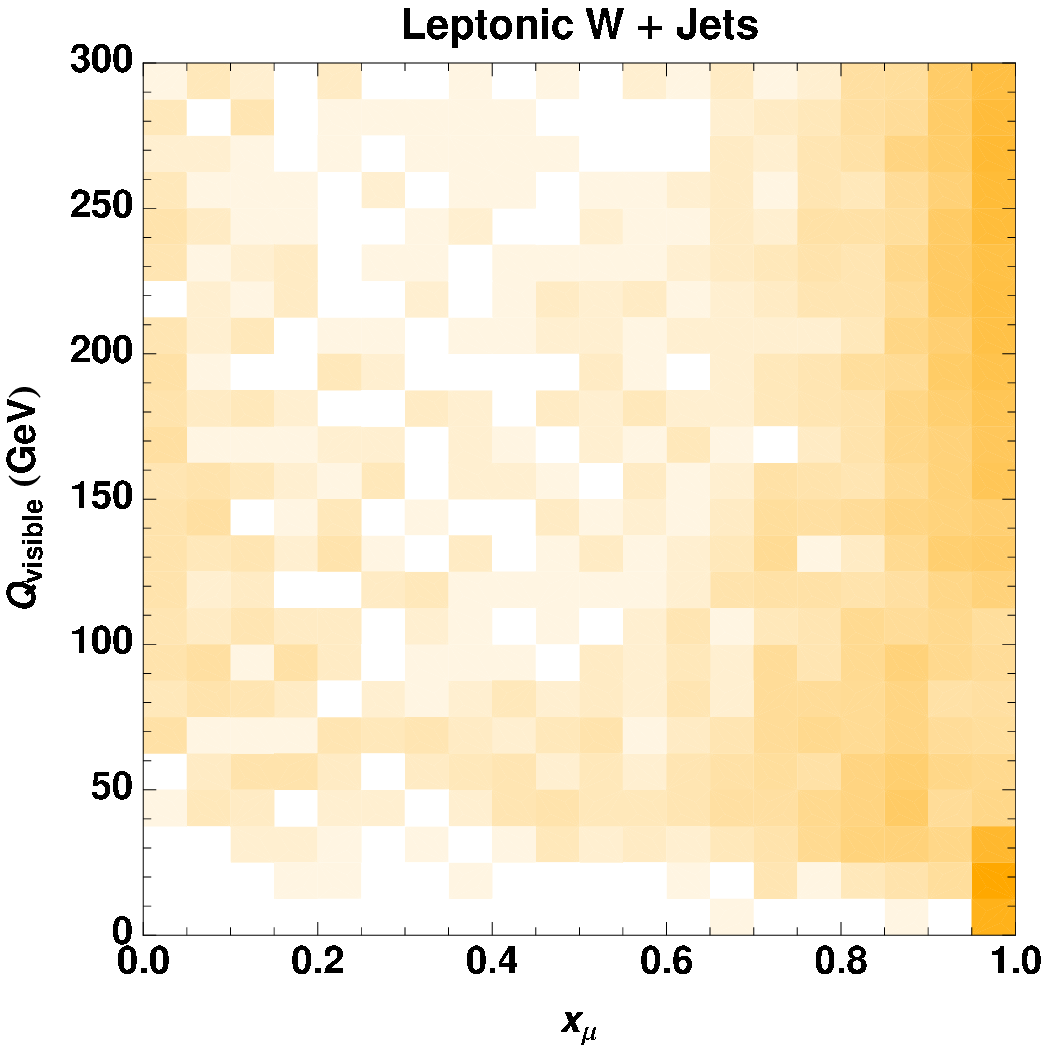} $\qquad$
\includegraphics[scale=0.6]{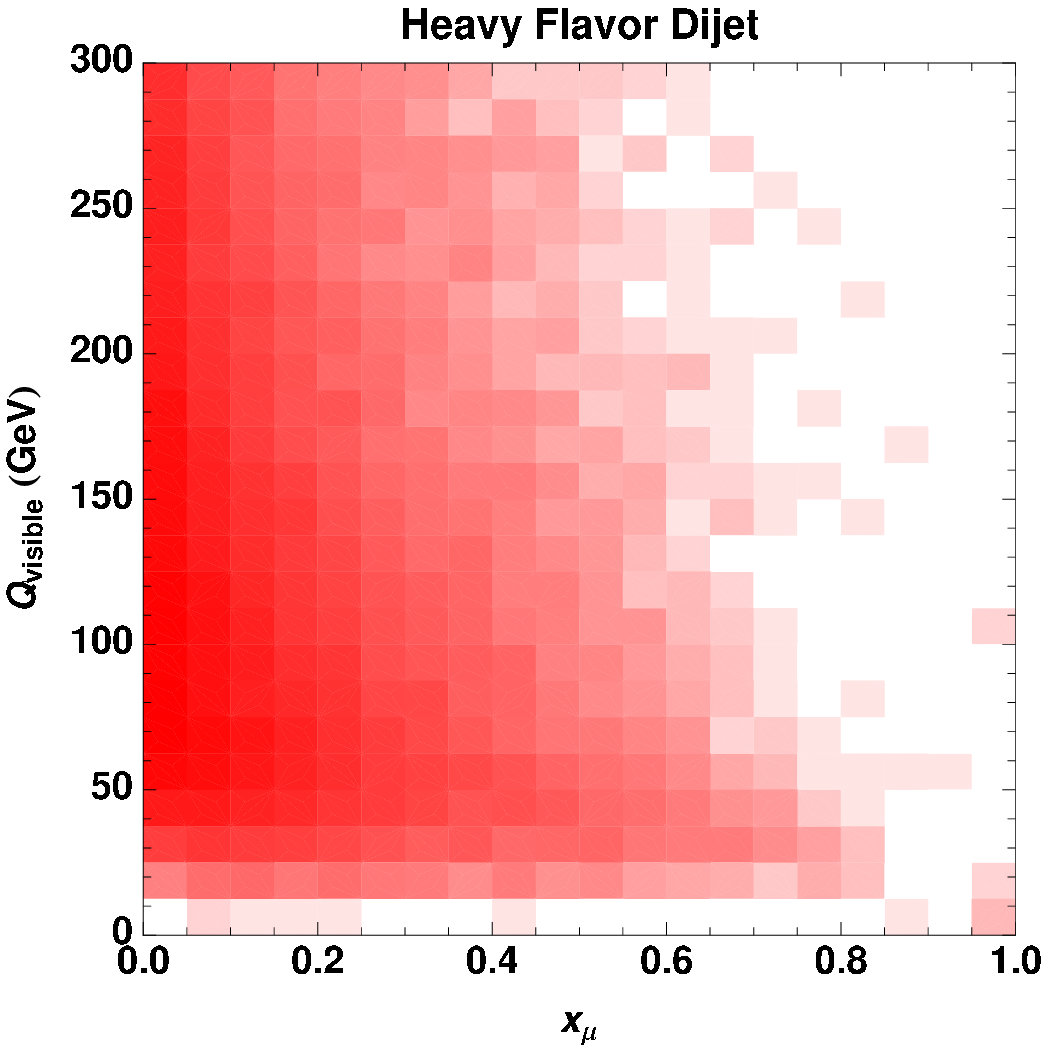} 
\end{center}
\caption{\label{fig:mu800_xl} Two dimensional distributions of
  $x_{\mu}$ (horizontal axis) versus visible mass $Q_{\rm visible}$
  (vertical axis) of the hardest muonic jet with a 800 GeV $p_T$ cut
  (including $p_T$ of muon).
  Top left:  leptonic top decays.  Because of the large neutrino
  missing energy, $Q_{\rm visible}$ peaks below $m_t$.  As explained
  in the text, $x_\mu$ is peaked near 1 because most of the visible
  mass comes from the leptonic-hadronic angular separation. 
Top right:  
hadronic top decays with leptonic $b$ decays.  Since there is little
missing energy in leptonic $b$ decays, there is a peak at $Q_{\rm
  visible} \sim m_t$.  As explained in the text, $x_\mu$ peaks near 0
because most of the visible mass comes from hadronic system alone. 
Bottom left:   
$W+\mbox{jets}$ production.  Just like the leptonic top decay, $x_\mu$ is peaked near 1.  
Bottom right:
Heavy flavor $b \bar{b}$ and $c \bar{c}$ production.  Just like the hadronic top decay, $x_\mu$ is peaked near 0.  
The intensity of the shading is proportional to the logarithm of the occupancy.
}
\end{figure}

For the boosted top signal, we will use the same $X$ resonances as \Sec{subsec:study-z}, with looser $p_T$ requirements to account for the missing neutrino:
\be
\label{eq:leptonicresonancemassandcut}
\begin{tabular}{c|c}
Resonance Mass & $p_T$ Cut \\
\hline
2 TeV & \phantom{0}600 GeV \\
3 TeV & \phantom{0}800 GeV \\
4 TeV & 1200 GeV
\end{tabular}
\ee
Using Monte Carlo truth information, we further divide the resonance
signal into a leptonic sample where the muon comes from $W$ decay and
a hadronic sample where the muon comes from leptonic $b$ decays.  We
use two different samples to estimate the irreducible backgrounds: 
\be
\label{eq:leptonicpythiaversions}
\begin{tabular}{l|l}
Name & Description \\
\hline
Heavy Flavor & \pythiaeightversion\ with $b\bar{b}$ and $c\bar{c}$ production \\
$W + \mbox{Jets}$ & \pythiaeightversion\ with $W + j$ production, forcing $W \rightarrow \mu \bar{\nu}_\mu$  \\
\end{tabular}
\ee
Because we only look at the hadronic activity surrounding the hardest
muon, the $W + \mbox{jets}$ sample effectively includes the same
physics as $Z + \mbox{jets}$.  Similarly, the heavy flavor sample
should give a reasonable approximation to any heavy flavor jet with
the $p_T$ cuts in \eq{eq:leptonicresonancemassandcut}. 

\begin{figure}
\begin{center}
\includegraphics[scale=0.6]{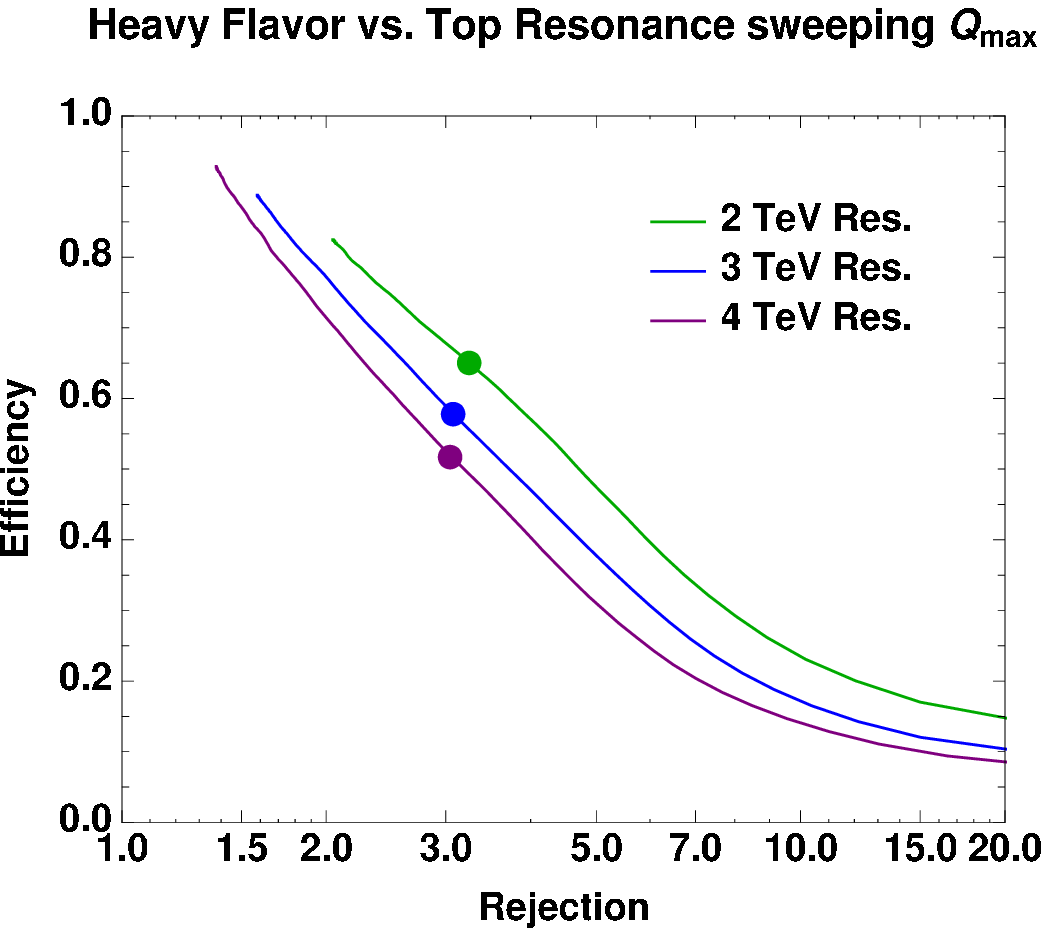} $\qquad$
\includegraphics[scale=0.6]{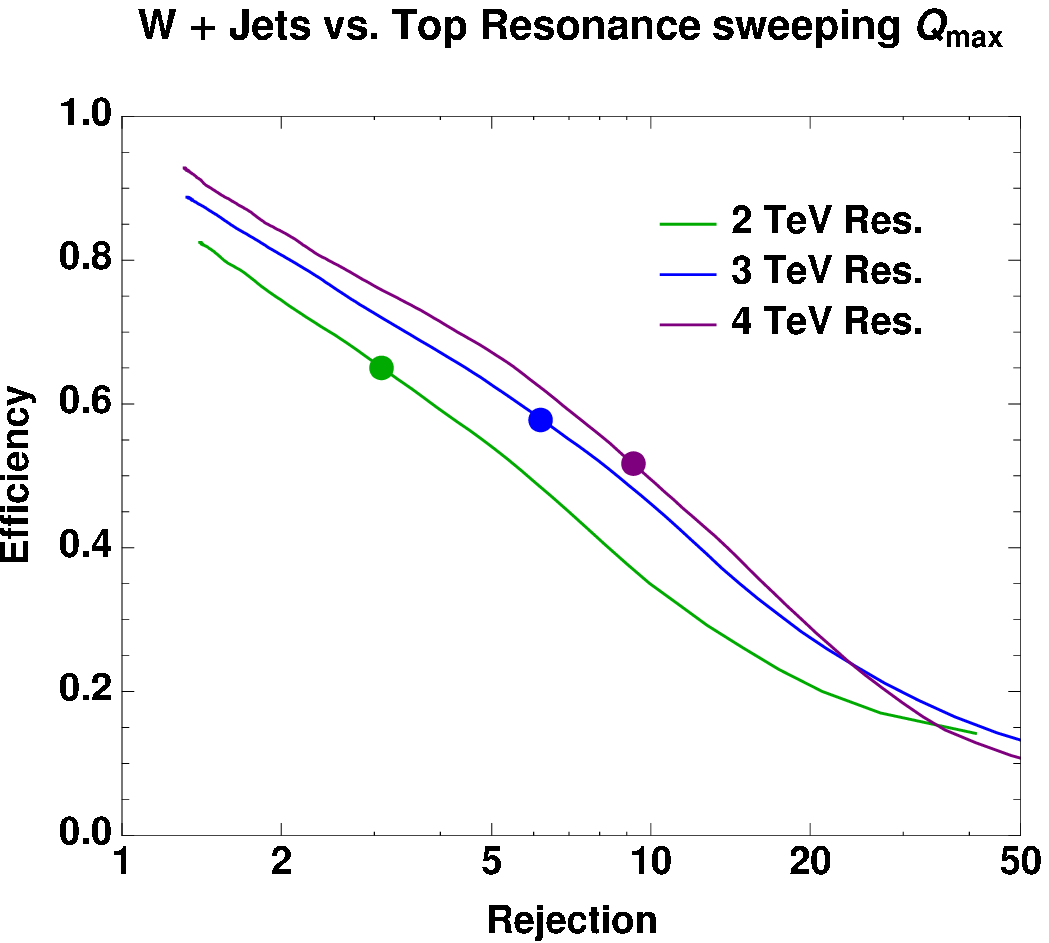}
\end{center}
\caption{\label{fig:invmasslepeffvsrej} Signal efficiency versus
  background rejection using a leptonic top mass window selection
  criteria $100 \GeV < Q_{\rm visible} < Q_{\rm max}$ with a varying
  $Q_{\rm max}$.  The backgrounds and signals are described in
  \eqs{eq:leptonicresonancemassandcut}{eq:leptonicpythiaversions}.
  Left:  heavy flavor background.  Right:  $W+\mbox{jets}$ background.
  The dots indicate the fiducial value $Q_{\rm max} = 200 \GeV$.
  Because the leptonic top has a wider variation in visible mass than
  the hadronic case, invariant mass alone is not as efficient at
  rejecting background.} 
\end{figure}

In \fig{fig:mu800_xl}, we show two-dimension distributions for
$\{Q_{\rm visible}, x_{\mu}\}$ for the 3 TeV top resonance and the two
background samples, all with an 800 GeV $p_T$ (including the $p_T$ of the
muon) cut on the muonic jet. 
We plot $x_{\mu}$ only for comparison purposes, and will
  discuss the $x_{\mu}$ spectrum in more detail below. In order to
isolate the boosted leptonic top 
sample, we want to impose a leptonic top mass window $Q_{\rm min} < 
Q_{\rm visible} < Q_{\rm max}$.  Since the heavy flavor dijet
background is peaked at low values of $Q_{\rm visible}$, we choose a
reasonably hard cut of $Q_{\rm min} = 100 \GeV$.  The effect of
changing $Q_{\rm max}$ on efficiency versus rejection is shown in
\fig{fig:invmasslepeffvsrej}, and we take $Q_{\rm max} = 200 \GeV$ as
a fiducial value. 

\begin{figure}
\begin{center}
\includegraphics[scale=0.6]{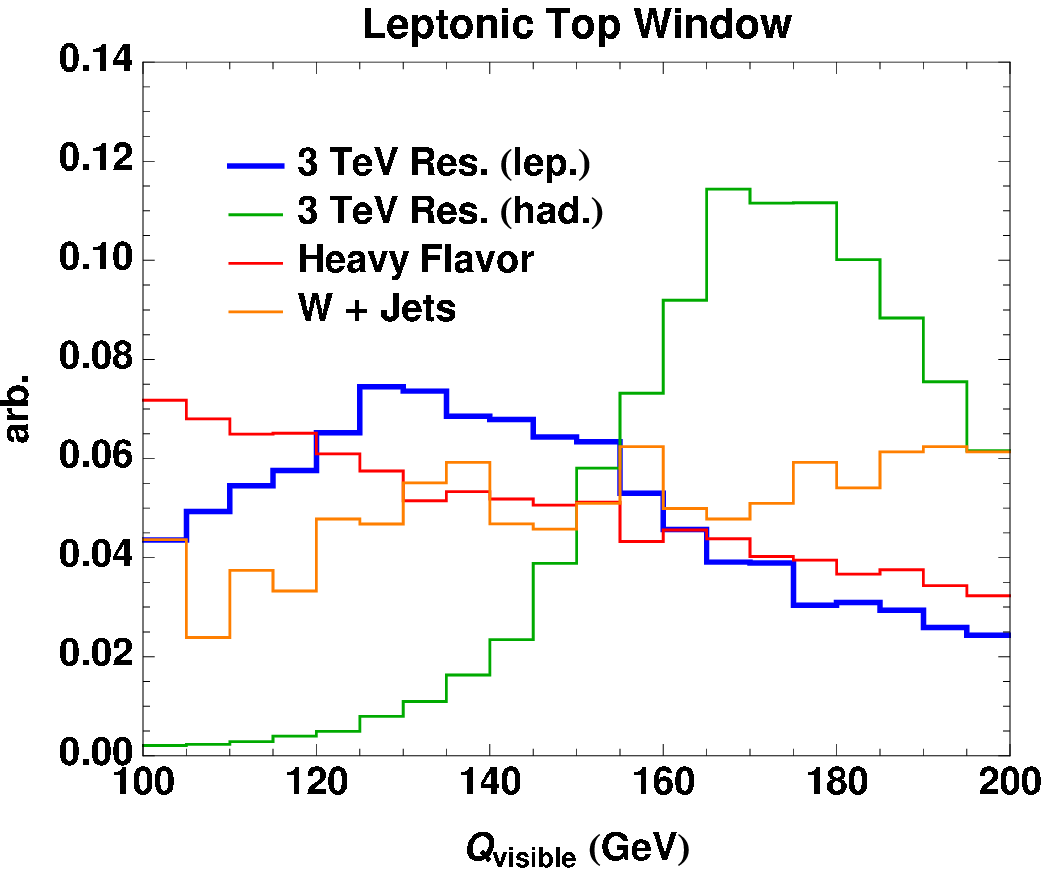} $\qquad$
\includegraphics[scale=0.6]{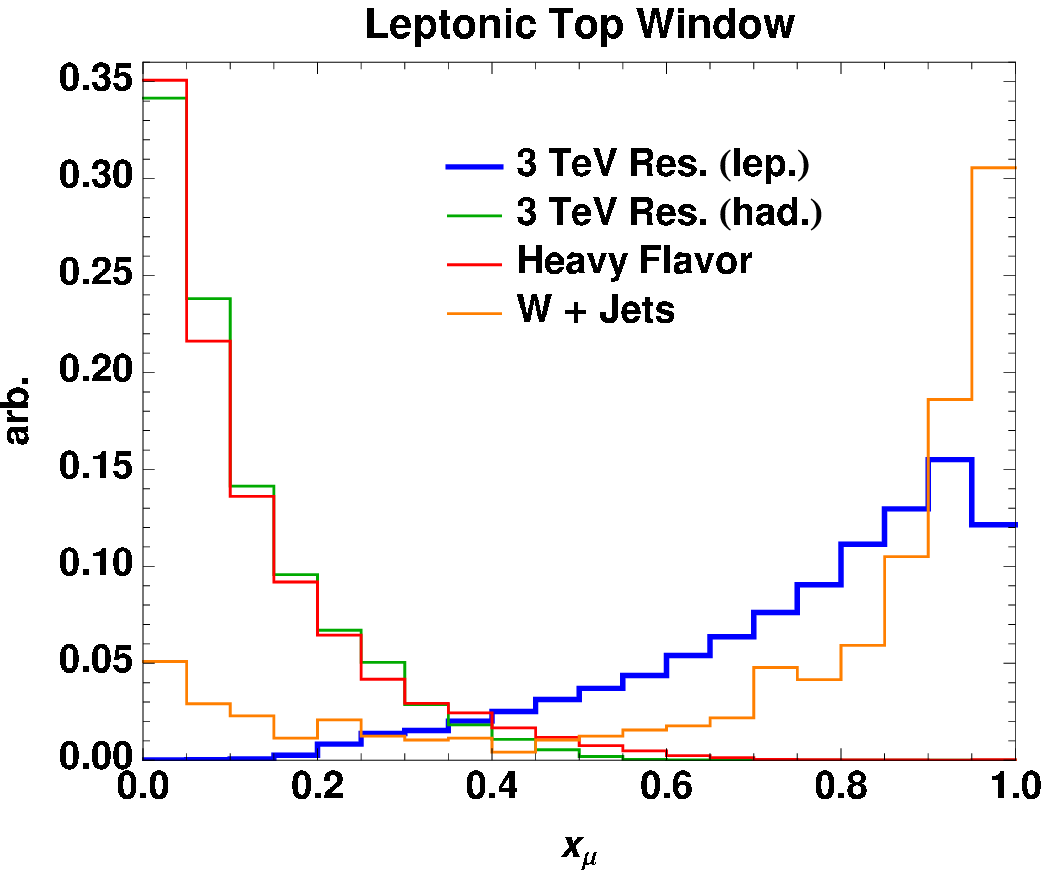}

~\\

\includegraphics[scale=0.6]{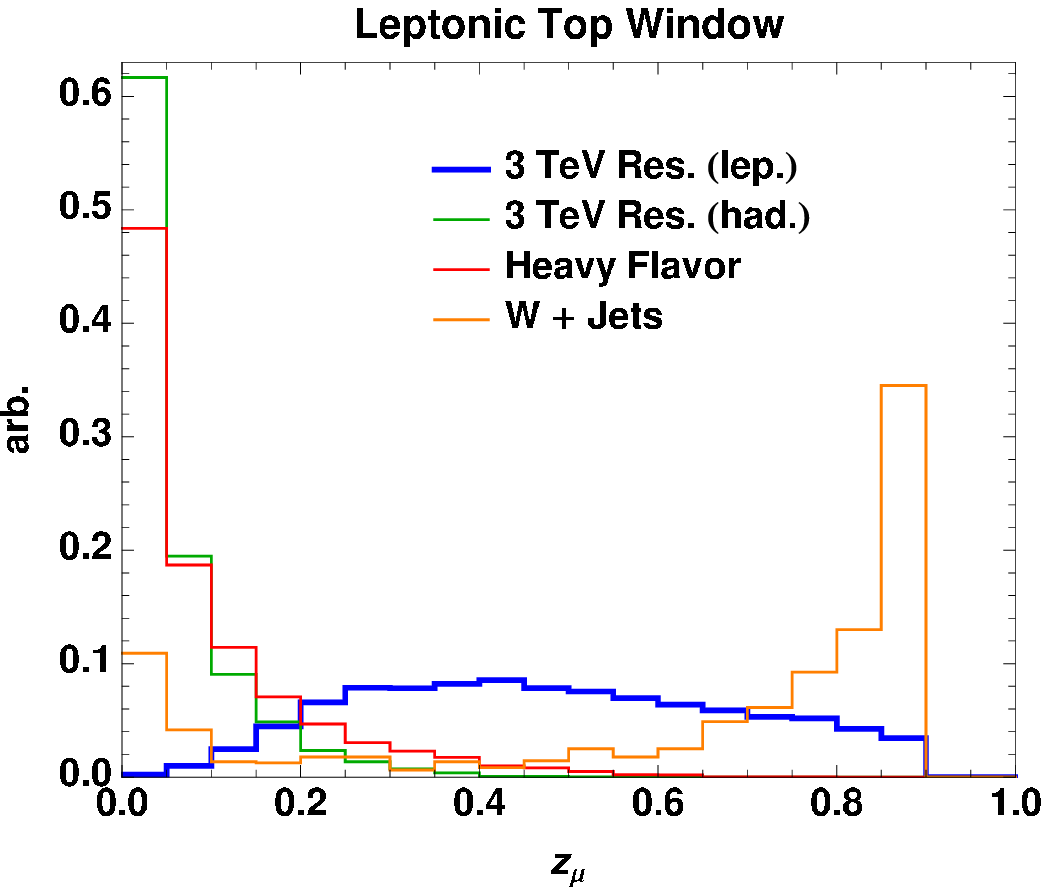} $\qquad$
\includegraphics[scale=0.6]{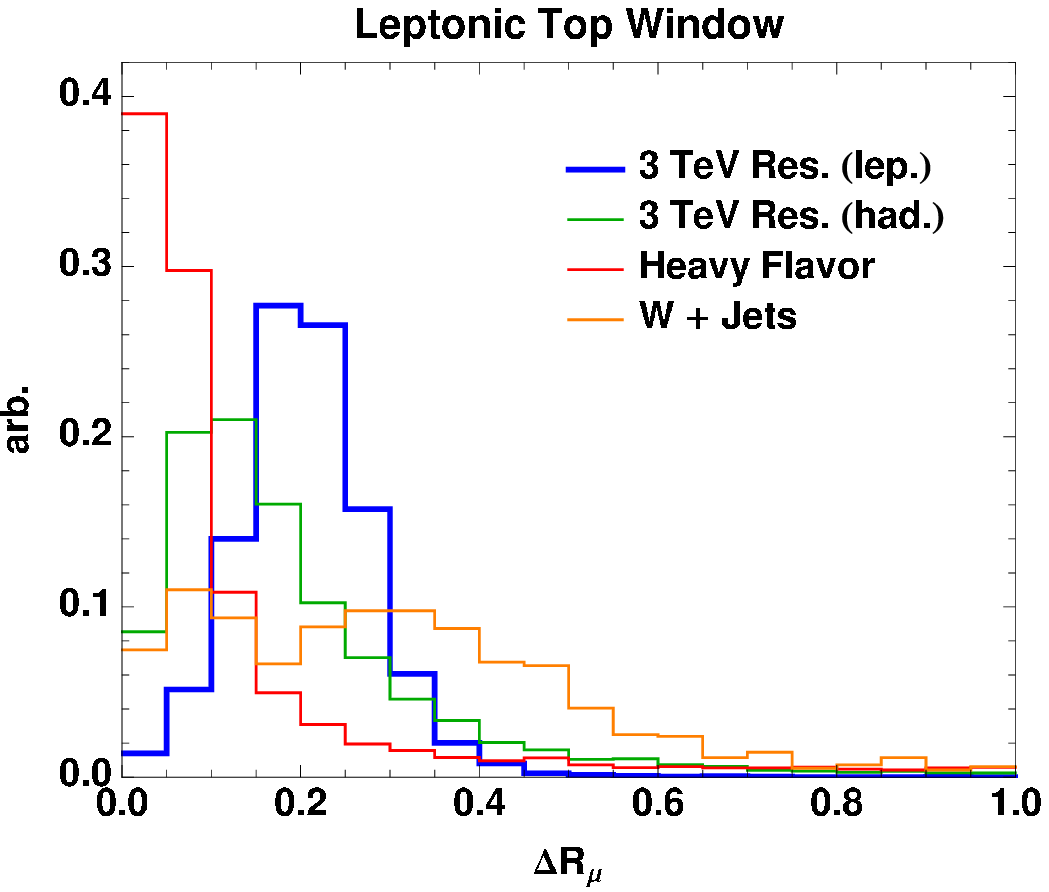} 
\end{center}
\caption{\label{fig:muon800} Muonic jet spectra of the hardest muonic
  jet after applying a leptonic top mass window $100 \GeV < Q_{\rm
    jet} < 200 \GeV$ cut, comparing leptonic tops (blue) with hadronic
  tops with leptonic $b$s (green), the heavy flavor background (red),
  and the $W+\mbox{jets}$ background (orange).  All samples have a jet
  $p_T$ cut of 800 GeV.  From left to right, top to bottom:  visible
  mass $Q_{\rm visible}$, $x_{\mu}$, $z_{\mu}$, and $\Delta R$.
  Descriptions of the behavior of these variables are given in the
  text. }
\end{figure}

Leptonic spectra for the 3 TeV resonance after imposing the $100 \GeV
< Q_{\rm visible} < 200 \GeV$ leptonic top window are shown in
\fig{fig:muon800}.  The simplest variable to understand is $z_\mu$. 
A lepton from a top decay is prompt and therefore carries a large
energy fraction of the 
top system.  A lepton from a bottom (charm) jet comes only from the
decay of a $B$ ($D$) meson, but before the meson is formed, a
significant fraction of the jet energy is already carried by radiated
gluons.  Therefore, $z_\mu$ is peaked closer to $0$ both for heavy
flavor jets and for hadronic tops with leptonic $b$ decays.  For the
$W + \mbox{jets}$ sample, $z_\mu$ peaks near 1 since most of the muons
are isolated, but some fraction of the time a jet from initial state
radiation can accidentally overlap with the muon.\footnote{There is a
  small peak near $z_\mu = 0$ for the $W + \mbox{jets}$ sample,
  because of $W + c$ production with leptonic $D$ decays.}  The
leptonic boosted tops occupies the middle range of $z_\mu$ and peaks
at 0.4, consistent with the expectation that the muon should carry
around 1/3 of the visible energy.\footnote{A better estimate
  accounting for the $W$ mass is $z^{\rm peak}_\mu = (m_t^2 +
  m_W^2)/(3 m_t^2 - m_W^2) \sim 0.44$.} 

The differences between the four samples are also seen in the angular
separation $\Delta R$ between the 
muon and the hadronic activity.  For heavy flavor jets, the muon is
aligned almost exactly along the jet direction, as expected since the
muon tracks the $B$ meson which tracks the $b$ quark.  For hadronic
tops with leptonic $b$ decays, there is an additional deflection in
$\Delta R$ from recoil against the $W$ boson.  The $\Delta R$ offset
grows larger for leptonic tops since the bulk of the energy follows
the hadronic activity.  For $W + \mbox{jets}$ the $\Delta R$ is
random, as there is no intrinsic correlation between the muon
direction and the hadronic activity.  

\begin{figure}[p]
\begin{center}
\includegraphics[scale=0.6]{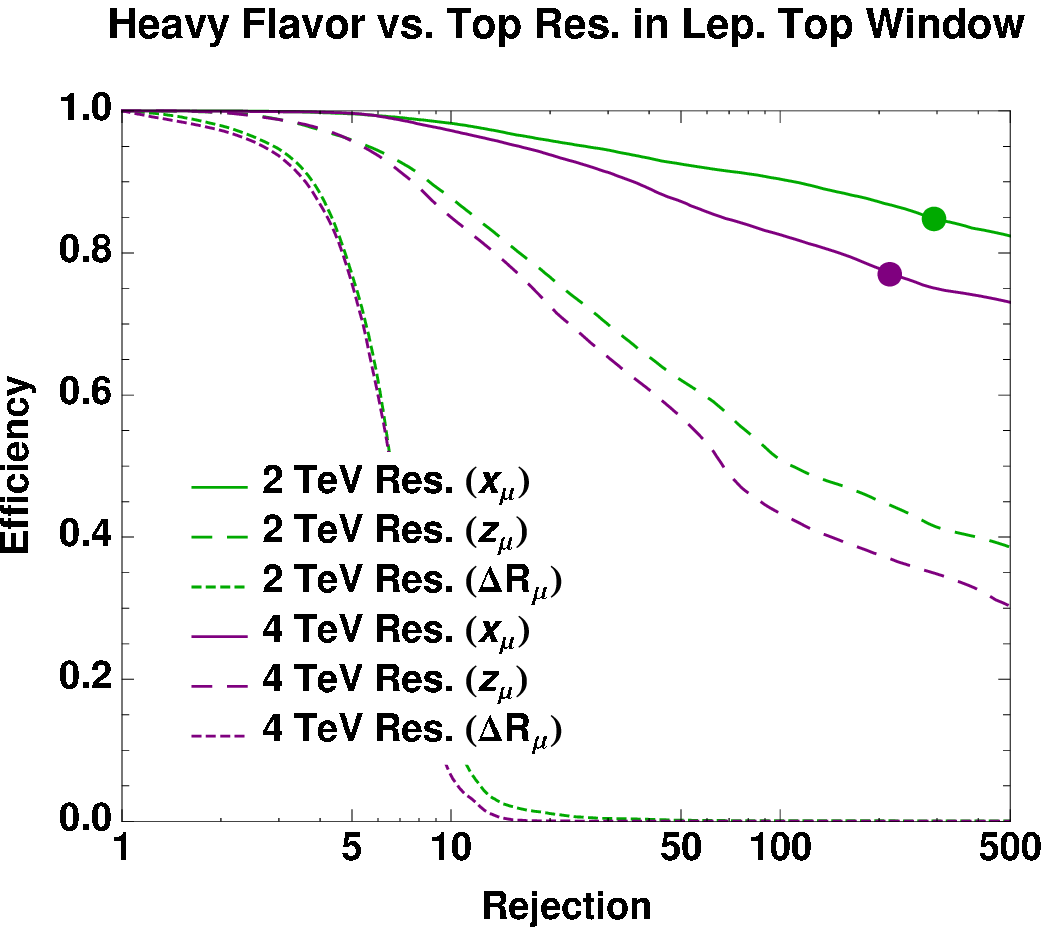}$\qquad$
\includegraphics[scale=0.6]{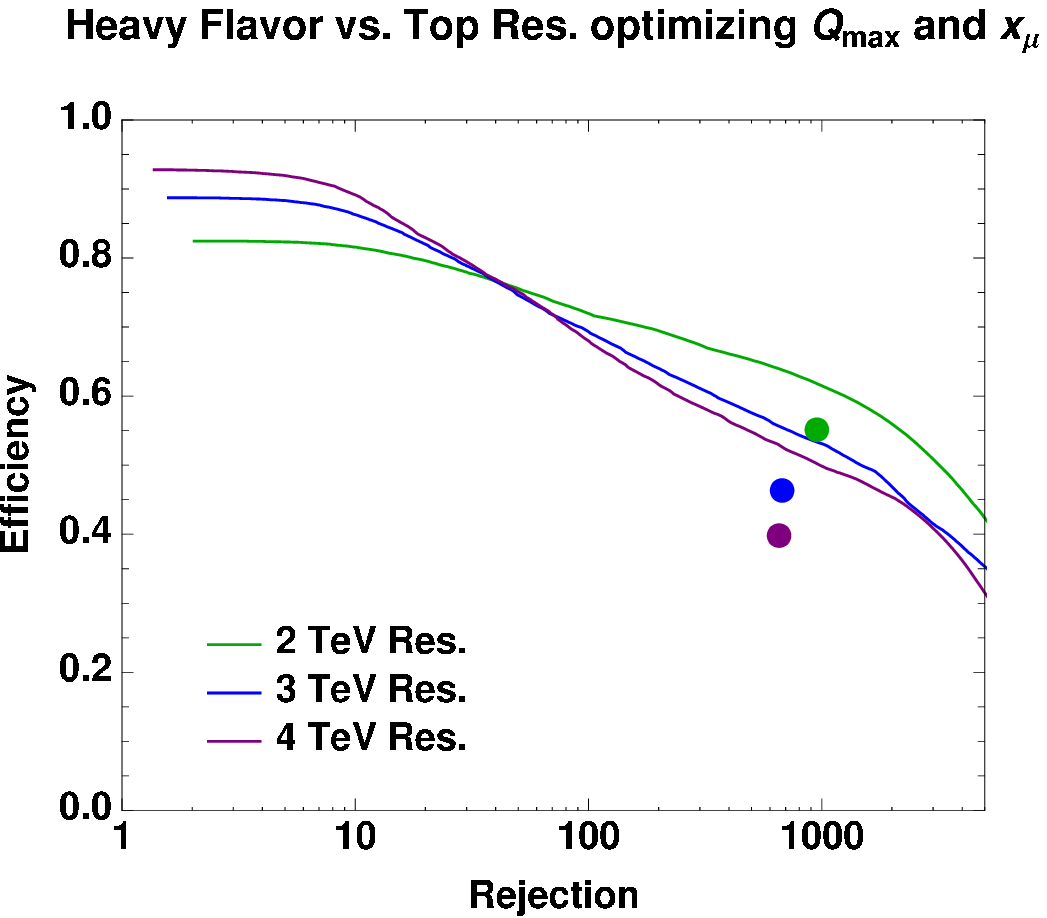} 

~\\

\includegraphics[scale=0.6]{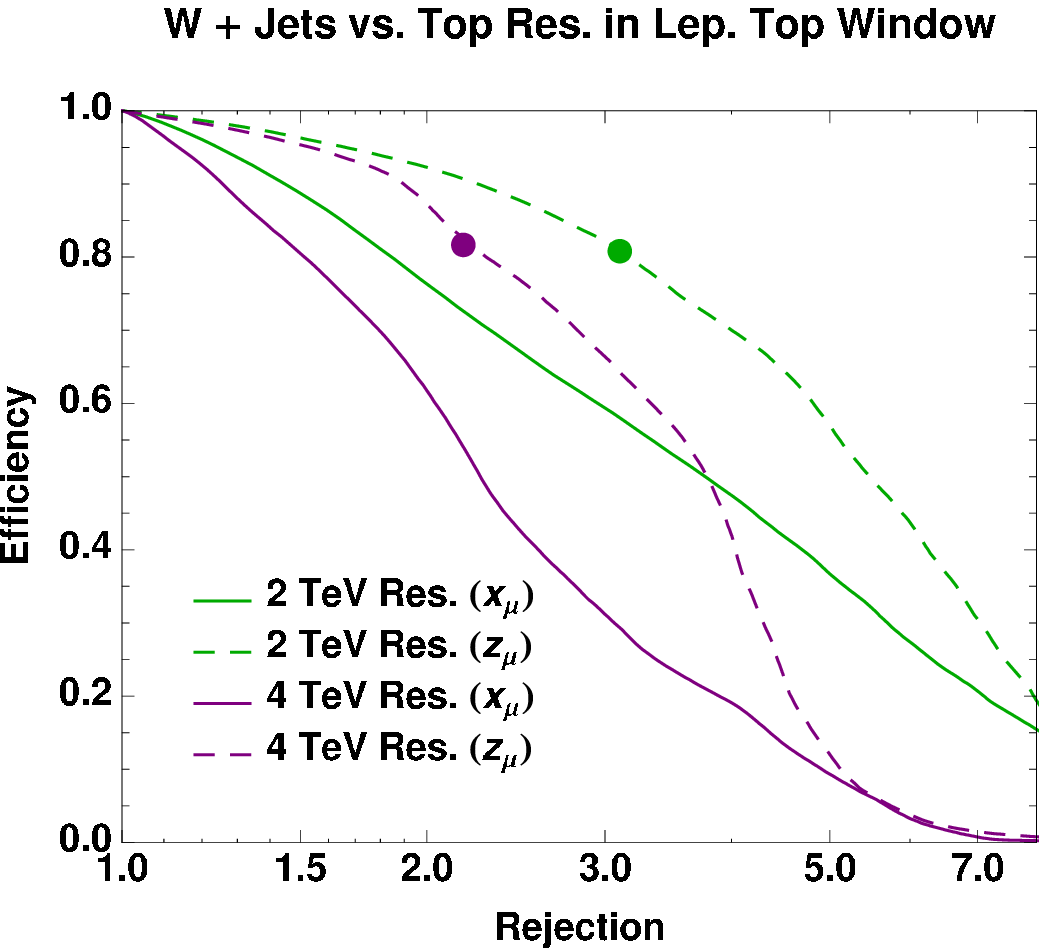}$\qquad$
\includegraphics[scale=0.6]{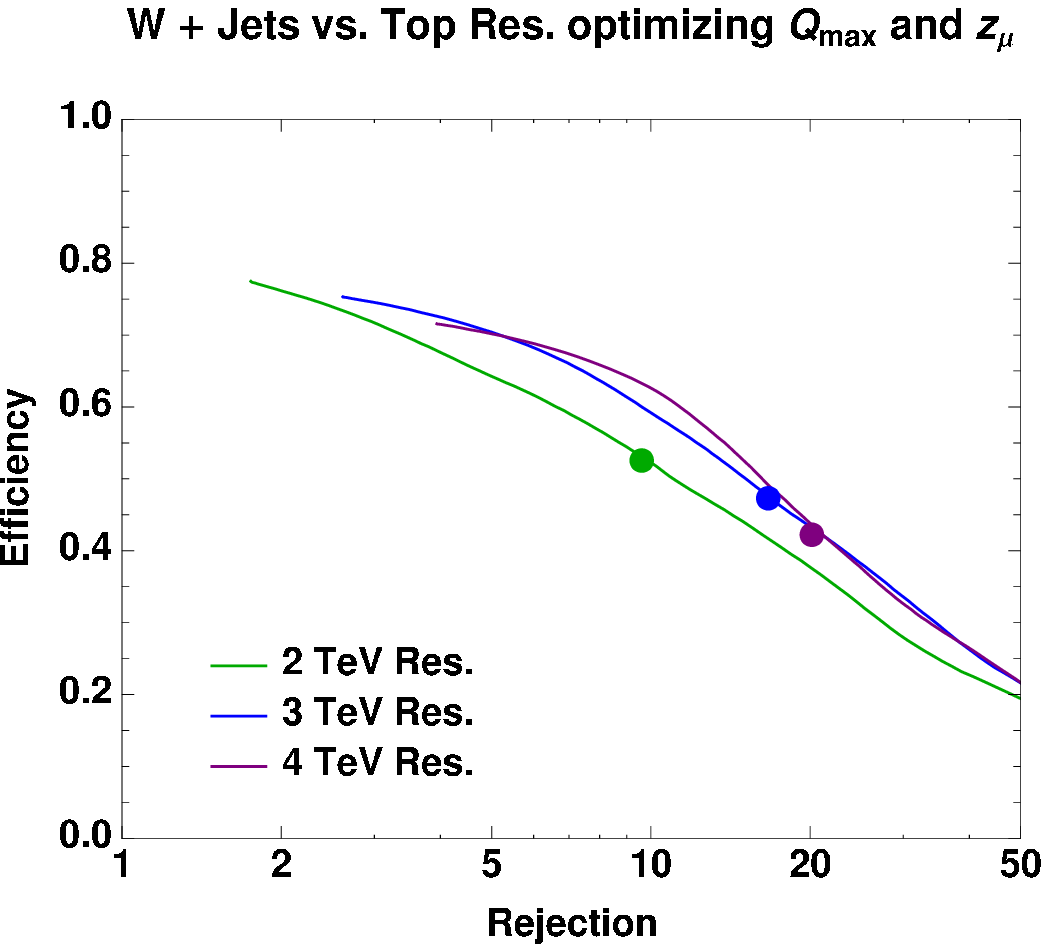}

\end{center}
\caption{\label{fig:eff-rej-xl}   Signal efficiency versus background
  rejection using the various leptonic variables. The top row is the
  heavy flavor background and the bottom row is the $W+\mbox{jets}$
  background.  The left plots are efficiency vs. rejection relative to
  the $100 \GeV < Q_{\rm visible} < 200 \GeV$ leptonic top window cut.   
The dots indicate the fiducial values $Q_{\rm max} = 200 \GeV$, $x_\mu
= 0.6$, and $z_{\mu} = 0.7$.  The plots  in the right column are
directly comparable 
to \fig{fig:invmasslepeffvsrej} and optimize $Q_{\rm max}$ with
respect to $x_\mu$ (for heavy flavor) or $z_\mu$ (for
$W+\mbox{jets}$). 
}
\end{figure}

Finally, a powerful variable for separating leptonic tops from heavy
flavor jets is $x_\mu$, which measures how much of the visible
invariant mass is carried by the hadronic activity.  Taking the muon
to be massless, we can rewrite $x_\mu$ as 
\be
1 - x_\mu = \frac{m_{b}^2}{Q_{\rm visible}^2}.
\ee
A boosted top has large $p_T$, but the invariant mass of the resulting
$b$ jet is expected to be around 10\% of $m_t$ since the available
phase space is only a fraction of the top mass after sharing energy
with the $W$ boson, therefore $x_\mu$ is peaked towards 1.
Similarly, $W + \mbox{jets}$ peaks at $x_\mu = 1$ because the muon
and the jet are uncorrelated and can be at large separation, meaning
that $2 p_{\mu}\cdot p_b$ tends to give the dominant contribution in
$(p_{\mu} + p_b)^2$.   For heavy flavor jets, the lepton carries a
small fraction of the energy, so in order for a heavy flavor jet to
have large invariant mass, it must come mostly from the hadronic
activity, yielding $x_\mu$ near zero.  Hadronic tops are nearly
identical to heavy flavor jets with the addition of extra hadronic
activity from the $W$ decay, so it also peaks at $x_\mu = 0$.   Of
course, to use $x_\mu$ effectively, one needs good momentum resolution
on the muon, which becomes increasingly difficult at higher $p_T$.

In Fig.~\ref{fig:eff-rej-xl}, we summarize the the boosted leptonic
top tagging efficiency versus background rejection using the three
variables $z_\mu$, $\Delta R$, and $x_\mu$.  Without taking into
account momentum resolution issues, it appears that the $x_\mu$
variable can be used to completely eliminate the heavy flavor
background.  Therefore, using a large $x_\mu$ cut along with
$b$-tagging should give a very clean boosted leptonic top signal,
because the $b$-tagging will help reduce the muon fake background.
For rejecting the $W + \mbox{jets}$ background, the $z_\mu$ variable
gives a factor of 2 to 4 rejection for only a small loss in
efficiency.  It will be interesting to see how $z_\mu$ and $x_\mu$
perform to reduce the muon fake background in the absence of
$b$-tagging. We also remark that the fraction of $W+$jets
  events in which the lepton actually overlaps with a jet is small
  since there is no obvious correlation between them.

\section{Conclusions and Outlook}

The top quark holds a unique position in both SM and BSM physics.  As
a SM resonance, it has a spectacular three-body decay with multiple
distinguishing features like the $b$-tagged jet and the on-shell
intermediate $W$ boson.  As a BSM focal point, it often has large
couplings to new resonances and might be copiously produced in new
physics signals.  Depending on the mass scales involved, though, top
quarks from BSM signals could be produced at large boosts, in which
case the three-body nature of the top quark decay is obscured.  While
the large invariant mass of the top quark helps separate boosted tops
from ordinary jets, fat jets are quite common at high $p_T$, therefore
it is important to find complementary variables to help increase the
purity of a boosted top signal.   

For boosted hadronic tops, we found that jet substructure observables
can give a factor of 2 to 5 better background rejection over invariant
mass alone with only a moderate loss of signal efficiency.  We
classified jet substructure variables into those directly related to a
$1 \rightarrow 2$ splitting process in the parton shower approximation
and those involving multiple objects.  Two-body observables are less
sensitive to the parton shower modeling, and the energy sharing
variable $z$ provides additional information about the identity of a
fat jet.  There is in principle more information in multi-body
signatures, and we found that $W$ reconstruction was a promising
strategy for boosted top identification, however more theoretical
studies are needed to understand multi-body jet substructure.  

For boosted leptonic tops, one can look for a hard muon surrounded by
hadronic activity, and we found that the spectrum of the muon relative
to the hadronic activity gave excellent rejection against the
irreducible backgrounds.  In particular, the heavy flavor background
could be almost completely eliminated through the $x_\mu$ variable,
and the $W/Z + \mbox{jets}$ background could be made manageable through
the $z_\mu$ variable.  However, further study is needed to understand
the issue of lepton fakes to test whether the spectrum of muons from
inflight decay or punch-through is sufficiently different from boosted
tops. 

In this study, we suggested variables based only on theoretical
considerations, and tested their effectiveness with minimal realism on
calorimeter segmentation.   In order to optimize these variables and
obtain realistic estimates of tagging efficiencies versus background
rejection, one needs to include more detector effects, such as
tracking information, energy/momentum resolution, and fakes.  Full
optimization of cuts should come after training these variables on
real jets at the LHC.  One important omission is that we did not make
use of $b$-tagging, which should help improve rejection against QCD
fat jets and lepton fakes, though the efficiency of $b$-tagging for
boosted tops at high $p_T$ is expected to decrease \cite{b-tag-hi-pt}.  

The results of this paper apply to the properties of individual fat
jets, ignoring any other features of the event.  In many BSM
scenarios, the signal is found not just with a single boosted top but
in a boosted $t \bar{t}$ channel.  Therefore, in addition to the
single top tagging observables we have studied in this paper,
variables which correlate the two top jets should be explored to
enhance the signal reach.

\section*{Acknowledgements}
We would like to thank Ben Lillie and Lisa Randall for collaboration
during the early stages of this project and for many valuable
discussions on jet substructures. We would also like to thank Aaron
Pierce and Jianmin Qian for discussions regarding leptonic variables.
J.T. is supported by a fellowship from the Miller Institute for Basic
Research in Science.   
The work of L.W.  is supported by the National Science Foundation
under Grant No.~0243680.

Shortly after this paper was posted, Ref.~\cite{Kaplan:2008ie} appeared which presented a complementary boosted top strategy.

\appendix

\section{The QCD Soft Singularity}
\label{app:qcdsoft}

In order to understand the QCD soft singularity, we have to understand
the expected $z$ distribution.  For illustrative purposes, we will
compare the $z$ distribution between a narrow resonance decay and QCD
radiation.  Because the decay products of a boosted top are
generically overlapping, we do not expect a boosted top to look
exactly like the narrow width expectation, but such a comparison is
still instructive.  To begin, we will describe two-body phase space
variables and then compare the differential distributions between the
two cases.

Consider the process $pp \rightarrow X A B$, where $A$ and $B$ are
subclusters of interest, and $X$ is every other final state in the
event.  We want to approximate the cross section for this process in a
factorized form as 
\be
\label{eq:XABfactorization}
d \sigma_{pp \rightarrow X A B} = d \sigma_{pp \rightarrow X M} d f_{M
  \rightarrow A B}, 
\ee
where the ``mother'' parton $M$ is a proxy for the fat jet.  If $M$ is
a resonance, then such a factorized form can be justified in the
narrow width approximation, with $d f_{M \rightarrow A B}$ being a
differential branching ratio.  This factorized form also makes sense
in the soft-collinear limit, where $d f_{M \rightarrow A B}$ is the
product of a differential splitting function \cite{ap} and a Sudakov factor \cite{Sudakov}. 

The phase space variables for the ``decay'' $M \rightarrow A B$ (with
masses $Q_M$, $Q_A$, and $Q_B$, respectively) are
\be
\frac{dQ_M^2}{2 \pi} d \Phi_2^{M \rightarrow A B},
\ee
where the on-shell two
body phase space is given by 
\be
\label{eq:lorentzinvarianttwobody}
d \Phi^{M \rightarrow AB}_2 = \frac{\lambda(Q_M^2; Q_A^2, Q_B^2)}{32
  \pi^2} d\Omega_2, \qquad d\Omega_2 \equiv d \cos\theta \,  d \phi, 
\ee
with the usual phase space factor 
\be
\lambda(Q_M^2; Q_A^2, Q_B^2) = \sqrt{\left(1 - \frac{Q_A^2}{Q_M^2} -
  \frac{Q_B^2}{Q_M^2} \right)^2 - 4 \frac{Q_A^2}{Q_M^2}
  \frac{Q_B^2}{Q_M^2}}. 
\ee
For unpolarized measurements, the $\phi$-dependence is uniform, so the
phase space of $M \rightarrow AB$ is characterized by two independent
quantities $Q_M$ and $\cos \theta$.  In order to study the QCD soft
singularity, the natural variables for fat jets are the invariant mass
$Q_M$ and some energy sharing variable $z = E_A/E_M$,\footnote{In the
  main body of the text, we define $z$ as $\min(E_A, E_B)/E_M$ since
  $A$ and $B$ are indistinguishable.  Here, $A$ and $B$ have
  meaningful quantum numbers, so it makes sense to talk about $z =
  E_A/E_M$.} and in 
general there will be a Jacobian $d \cos \theta / dz$ in the
transformation from $\cos \theta$ to $z$. 

If $z$ is interpreted strictly as $E_A/E_M$, then two-body
  kinematics restricts the range for $z$ to be  
\be
\label{eq:zlimits}
\left| z -\frac{1}{2}\left( 1 + \frac{Q_A^2}{Q_M^2} -
\frac{Q_B^2}{Q^2_M} \right) \right| \leq \frac{\beta_M}{2}
\lambda(Q_M^2,Q_A^2,Q_B^2), \qquad \beta_M = \sqrt{1-Q_M^2/E_M^2},
\ee
where $\beta_M$ is the boost magnitude from the $M$ center-of-mass
frame to the lab 
frame.   Because these limits depend on $Q_A$ and $Q_B$, in a parton
shower with multiple emissions, the correct $z$ limits on $M
\rightarrow AB$ can only be determined after one knows how $A$ and $B$
will split which sets the values of $Q_A$ and $Q_B$.  In particular, a
value of $z$ that  satisfies 
\eq{eq:zlimits} for $Q_{A,B} = 0$ might be invalid for $Q_{A,B}>0$.
There are various ways to deal with this ambiguity \cite{reshuffle},
and most parton showers employ some kind of momentum reshuffling
procedure, but it means that the interpretation of $z$ in $d f_{M
  \rightarrow A B}$ can depend on $Q_A$ and $Q_B$ in a non-trivial and
algorithm-specific way.

We can now compare the differential distributions $d f_{M \rightarrow
  A B}$ between the narrow width approximation and QCD radiation.  In
the narrow width approximation, the mother $M$ is exactly on-shell: 
\be
d f^{\rm NWA}_{M \rightarrow A B} = \frac{dQ_M^2}{2 \pi} \frac{d
  \Phi_2^{M \rightarrow A B}}{V_2} \Br(M \rightarrow A B) \delta(Q_M^2
- m_M^2), 
\ee
where $V_2 \equiv \int d\Phi_2$ is the volume of two-body Lorentz
invariant phase space, which depends on the masses of $A$ and $B$.
Note that $d f^{\rm NWA}$ is uniform in $\cos \theta$. 

In the soft-collinear QCD case, one can use a parton shower language \cite{reshuffle}
where the natural variables are the evolution variable $\mu$ and the
energy sharing variable $z$, both of which are functions of $\{Q_M^2,
\cos \theta\}$.  Unlike the narrow width approximation, the parton $M$
is never on-shell, and its off-shellness is determined by the
evolution variable $\mu(Q_M^2,\cos \theta)$.  Using unpolarized
splitting functions defined in terms of the energy sharing variable
$z(Q_M^2, \cos \theta)$, the QCD splitting is described by 
\be
d f^{\rm QCD}_{M \rightarrow A B} = d \log \mu^2\,
\frac{d\phi}{2\pi}\,  dz \,  \frac{\alpha_s(\mu)}{2\pi} P_{M
  \rightarrow A B}(z) \Delta(\mu_{\rm start}, \mu), 
\ee
where $P_{M \rightarrow A B}(z)$ are the usual Altarelli-Parisi
splitting functions \cite{ap}
\begin{align}
P_{q \to qg}(z) &= C_F\,\frac{1+z^2}{1-z}
,\nonumber\\
P_{g \to gg}(z) &= C_A\biggl[\frac{1-z}{z} + \frac{z}{1-z} + z(1-z)\biggr]
, \nonumber\\
P_{g \to q\bar{q}}(z) &= T_R\,\bigl[z^2+(1-z)^2\bigr],
\end{align}
and $\Delta(\mu_{\rm start}, \mu)$ is a Sudakov factor \cite{Sudakov}
\be
\Delta(\mu_{\rm start}, \mu) = \exp \left[-\sum_{AB}
  \int_\mu^{\mu_{\rm start}} d \log \mu' \int \frac{d\phi}{2\pi} \int  dz \,
  \frac{\alpha_s(\mu')}{2\pi} P_{M \rightarrow A B}(z)    \right]. 
\ee
Note the soft singularities at $z \rightarrow 0$ or $z \rightarrow 1$
which are equivalent to singularities at $|\cos \theta| \rightarrow
1$.  

We want to describe both $d f^{\rm NWA}$ and $d f^{\rm QCD}$ in terms
of the natural boosted top variables $Q_M^2$ and $z$.  We have not yet
specified the functional form of $\mu$ and $z$, and in general, there
will be a non-trivial Jacobian in the transformation between $\{\mu,
z\}$ and $\{Q^2,\cos \theta\}$.  Typical evolution variables are
proportional to invariant mass as $\mu^2 = Q_M^2 f(z)$, in which case
$d \log\mu^2 = d \log Q_M^2$.  Integrating over $\phi$ for simplicity 
\begin{align}
d f^{\rm NWA}_{M \rightarrow A B} &= \frac{dQ_M^2}{2 \pi} d z  \frac{d
  \cos \theta}{dz} \Br(M \rightarrow A B) \delta(Q_M^2 - m_M^2), \\ 
d f^{\rm QCD}_{M \rightarrow A B} &= \frac{dQ_M^2}{2 \pi} d z
\frac{1}{Q_M^2} \frac{\alpha_s(\mu)}{(2\pi)} P_{M \rightarrow A B}(z)
\Delta(\mu_{\rm start}, \mu), \label{eq:realsplitqcd} 
\end{align}
where $\mu$ is still a function of $Q^2$ and $z$.  Assuming that $d
\cos \theta / dz$ does not contain any singularities, this function
has the form advertised in \eqs{eq:simpsplitnwa}{eq:simpsplitqcd}.  If
$Q_A = Q_B = 0$, then $d \cos \theta / dz = 2/\beta_M$.

We see that $d f^{\rm QCD}$ does not factorize into a
$Q_M^2$-dependent and a $z$-dependent piece, because the $\alpha_s$
factor and Sudakov factor are written in terms of the evolution
variable $\mu$ that mixes $Q^2$ and $z$.  On the other hand, at the
double logarithmic level the value $\alpha_s$ is fixed, and
because the definition of $\mu_{\rm start}$ is ambiguous in 
the Sudakov factor, the value of $\mu_{\rm start}$ can be chosen
remove the $z$-dependence up to subleading corrections.   
In other words, the non-factorizing pieces occur only at
the single logarithmic level, so $d f^{\rm QCD}$ does
factorize in the leading logarithmic approximation.   

To design the perfect $z$-variable for boosted tops, one would want to
figure out which 
$z$-like variable is maximally orthogonal to $Q^2$ for real QCD in the
neighborhood of $Q^2 \sim m_t^2$.  This requires understanding the
spectrum of $Q_A$ and $Q_B$ from subsequent splittings and how they
influence the interpretation of $z$.  Whether this is in principle
possible depends on whether the leading corrections to $d f^{\rm QCD}$ 
still respect the factorized form of \eq{eq:XABfactorization}.  If
higher order QCD corrections simply modify the functional form of  $d
f^{\rm QCD}$, then there is always some variable $z$ that is
orthogonal to $Q^2$, though in practice it may be difficult to find an
experimental observable correlated with $z$.  But if higher order QCD
corrections imply that the ideal choice of $z$ variable strongly
depends on the spectators $X$ in the process $pp \rightarrow X A B$,
then there is no way to improve the definition of $z$.  A
plausible intermediate possibility is that $d f^{\rm QCD}$ depends on
$X$ only in some limited way such as in the definition of $\mu_{\rm
  start}$.  If $z$-like variables are found to be useful in an
experimental context, then a detailed study of $d f^{\rm QCD}$ is
warranted.


\begin{thebibliography}{99}

\bibitem{cdf-top}
\verb$http://www-cdf.fnal.gov/physics/new/top/top.html$

\bibitem{d0-top}
\verb$http://www-d0.fnal.gov/Run2Physics/WWW/results/top.htm$

\bibitem{stt_lhc}
  R.~Bonciani, S.~Catani, M.~L.~Mangano and P.~Nason,
  Nucl.\ Phys.\  B {\bf 529}, 424 (1998)
  [arXiv:hep-ph/9801375].
For recent updates, see 
  S.~Moch and P.~Uwer,
  arXiv:0804.1476 [hep-ph];
  M.~Cacciari, S.~Frixione, M.~M.~Mangano, P.~Nason and G.~Ridolfi,
  arXiv:0804.2800 [hep-ph];
 N.~Kidonakis and R.~Vogt,
  arXiv:0805.3844 [hep-ph].


\bibitem{atlasTDR}
 ``ATLAS: Detector and physics performance technical design report. Volume
 1,''
CERN-LHCC-99-14. \\
``ATLAS detector and physics performance. Technical design report.  Vol. 2,''
CERN-LHCC-99-15
\bibitem{cmsTDR}
CMS Techinical Design Report V.1., CERN-LHCC-2006-001, \\
CMS Techinical Design Report V.2., CERN-LHCC-2006-021


\bibitem{Dimopoulos:1981zb}
 S.~Dimopoulos and H.~Georgi,
 Nucl.\ Phys.\ B {\bf 193}, 150 (1981).
\bibitem{UED}
 T.~Appelquist, H.~C.~Cheng and B.~A.~Dobrescu,
 Phys.\ Rev.\  D {\bf 64}, 035002 (2001)
 [arXiv:hep-ph/0012100].
\bibitem{ArkaniHamed:2001nc}
 N.~Arkani-Hamed, A.~G.~Cohen and H.~Georgi,
 Phys.\ Lett.\  B {\bf 513}, 232 (2001)
 [arXiv:hep-ph/0105239].
\bibitem{twinhiggs}
 Z.~Chacko, H.~S.~Goh and R.~Harnik,
 Phys.\ Rev.\ Lett.\  {\bf 96}, 231802 (2006)
 [arXiv:hep-ph/0506256].


\bibitem{tprime-model}
 H.~C.~Cheng, I.~Low and L.~T.~Wang,
 Phys.\ Rev.\ D {\bf 74}, 055001 (2006)
 [arXiv:hep-ph/0510225].
\bibitem{tprime-study-1}
 P.~Meade and M.~Reece,
 Phys.\ Rev.\ D {\bf 74}, 015010 (2006)
 [arXiv:hep-ph/0601124].

\bibitem{tprime-study-2}
 A.~Freitas and D.~Wyler,
 JHEP {\bf 0611}, 061 (2006)
 [arXiv:hep-ph/0609103].
   A.~Belyaev, C.~R.~Chen, K.~Tobe and C.~P.~Yuan,
 arXiv:hep-ph/0609179.

\bibitem{Nojiri}
 S.~Matsumoto, M.~M.~Nojiri and D.~Nomura,
 Phys.\ Rev.\  D {\bf 75}, 055006 (2007)
 [arXiv:hep-ph/0612249];
 M.~M.~Nojiri and M.~Takeuchi,
 arXiv:0802.4142 [hep-ph].


\bibitem{top-composite}
 T.~Gherghetta and A.~Pomarol,
 Nucl.\ Phys.\  B {\bf 586}, 141 (2000)
 [arXiv:hep-ph/0003129].
 K.~Agashe, A.~Delgado, M.~J.~May and R.~Sundrum,
 JHEP {\bf 0308}, 050 (2003)
 [arXiv:hep-ph/0308036].

\bibitem{resonance-ttbar}  
K.~Agashe, A.~Belyaev, T.~Krupovnickas, G.~Perez and J.~Virzi,
 Phys.\ Rev.\  D {\bf 77}, 015003 (2008)
 [arXiv:hep-ph/0612015].
V.~Barger, T.~Han and D.~G.~E.~Walker,
 Phys.\ Rev.\ Lett.\  {\bf 100}, 031801 (2008)
 [arXiv:hep-ph/0612016].
 B.~Lillie, L.~Randall and L.~T.~Wang,
 JHEP {\bf 0709}, 074 (2007)
 [arXiv:hep-ph/0701166].
 A.~L.~Fitzpatrick, J.~Kaplan, L.~Randall and L.~T.~Wang,
 JHEP {\bf 0709}, 013 (2007)
 [arXiv:hep-ph/0701150].
 K.~Agashe, H.~Davoudiasl, G.~Perez and A.~Soni,
 Phys.\ Rev.\  D {\bf 76}, 036006 (2007)
 [arXiv:hep-ph/0701186].
 U.~Baur and L.~H.~Orr,
 Phys.\ Rev.\  D {\bf 76}, 094012 (2007)
 [arXiv:0707.2066 [hep-ph]].
 R.~Frederix and F.~Maltoni,
 arXiv:0712.2355 [hep-ph].
 U.~Baur and L.~H.~Orr,
 arXiv:0803.1160 [hep-ph].



\bibitem{top_convention}
 B.~Abbott {\it et al.}  [D0 Collaboration],
 Phys.\ Rev.\ D {\bf 58}, 052001 (1998)
 [arXiv:hep-ex/9801025].
 F.~Abe {\it et al.}  [CDF Collaboration],
 Phys.\ Rev.\ Lett.\  {\bf 80}, 2767 (1998)
 [arXiv:hep-ex/9801014].


\bibitem{pythia}
 T.~Sjostrand, S.~Mrenna and P.~Skands,
 JHEP {\bf 0605}, 026 (2006)
 [arXiv:hep-ph/0603175].
  T.~Sjostrand, S.~Mrenna and P.~Skands,
  arXiv:0710.3820 [hep-ph].

\bibitem{fastjet}
  M.~Cacciari and G.~P.~Salam,
  Phys.\ Lett.\  B {\bf 641}, 57 (2006)
  [arXiv:hep-ph/0512210], 
\verb$http://www.lpthe.jussieu.fr/~salam/fastjet/$.

\bibitem{anti-kt}
 M.~Cacciari, G.~P.~Salam and G.~Soyez,
 JHEP {\bf 0804}, 063 (2008)
 [arXiv:0802.1189 [hep-ph]].


\bibitem{ap}
  G.~Altarelli and G.~Parisi,
  Nucl.\ Phys.\  B {\bf 126}, 298 (1977).

\bibitem{Brooijmans}
G. Brooijmans, ATLAS note, ATL-PHYS-CONF-2008-008.

\bibitem{topmass}
L.~Almeida, S.~J.~Lee, G.~Perez,  G.~Sterman, I.~Sung, J.~S.~Virzi, in
preparation. 

\bibitem{reshuffle}
  T.~Sjostrand, P.~Eden, C.~Friberg, L.~Lonnblad, G.~Miu, S.~Mrenna
  and E.~Norrbin, 
  Comput.\ Phys.\ Commun.\  {\bf 135}, 238 (2001)
  [arXiv:hep-ph/0010017].
  T.~Sjostrand and P.~Z.~Skands,
  Eur.\ Phys.\ J.\  C {\bf 39}, 129 (2005)
  [arXiv:hep-ph/0408302].
  G.~Corcella {\it et al.},
  JHEP {\bf 0101}, 010 (2001)
  [arXiv:hep-ph/0011363].
  R.~Kuhn, F.~Krauss, B.~Ivanyi and G.~Soff,
  Comput.\ Phys.\ Commun.\  {\bf 134}, 223 (2001)
  [arXiv:hep-ph/0004270].
  F.~Krauss, A.~Schalicke and G.~Soff,
  Comput.\ Phys.\ Commun.\  {\bf 174}, 876 (2006)
  [arXiv:hep-ph/0503087].
  G.~Gustafson and U.~Pettersson,
  Nucl.\ Phys.\  B {\bf 306}, 746 (1988).
  L.~Lonnblad,
  Comput.\ Phys.\ Commun.\  {\bf 71}, 15 (1992).
  S.~Gieseke, P.~Stephens and B.~Webber,
  JHEP {\bf 0312}, 045 (2003)
  [arXiv:hep-ph/0310083].
  S.~Gieseke, A.~Ribon, M.~H.~Seymour, P.~Stephens and B.~Webber,
  JHEP {\bf 0402}, 005 (2004)
  [arXiv:hep-ph/0311208].
  S.~Gieseke {\it et al.},
  arXiv:hep-ph/0609306.
  C.~W.~Bauer and F.~J.~Tackmann,
  Phys.\ Rev.\  D {\bf 76}, 114017 (2007)
  [arXiv:0705.1719 [hep-ph]].
  C.~W.~Bauer, F.~J.~Tackmann and J.~Thaler,
  arXiv:0801.4028 [hep-ph].
  
  
\bibitem{jade}
  W.~Bartel {\it et al.}  [JADE Collaboration],
  Z.\ Phys.\  C {\bf 33}, 23 (1986).

\bibitem{durham}
  S.~Catani, Y.~L.~Dokshitzer, M.~Olsson, G.~Turnock and B.~R.~Webber,
  Phys.\ Lett.\  B {\bf 269}, 432 (1991).
  
\bibitem{geneva}
  S.~Bethke, Z.~Kunszt, D.~E.~Soper and W.~J.~Stirling,
  Nucl.\ Phys.\  B {\bf 370}, 310 (1992)
  [Erratum-ibid.\  B {\bf 523}, 681 (1998)].

\bibitem{kt}
  S.~Catani, Y.~L.~Dokshitzer, M.~H.~Seymour and B.~R.~Webber,
  Nucl.\ Phys.\  B {\bf 406}, 187 (1993).
  S.~D.~Ellis and D.~E.~Soper,
  Phys.\ Rev.\  D {\bf 48}, 3160 (1993)
  [arXiv:hep-ph/9305266].

\bibitem{Cambridge-Aachen}
  Y.~L.~Dokshitzer, G.~D.~Leder, S.~Moretti and B.~R.~Webber,
  JHEP {\bf 9708}, 001 (1997)
  [arXiv:hep-ph/9707323].
  G.~P.~Salam and G.~Soyez,
  JHEP {\bf 0705}, 086 (2007)
  [arXiv:0704.0292 [hep-ph]].


\bibitem{Wjmass}
 J.~M.~Butterworth, B.~E.~Cox and J.~R.~Forshaw,
 Phys.\ Rev.\ D {\bf 65}, 096014 (2002)
 [arXiv:hep-ph/0201098].
  J.~M.~Butterworth, A.~R.~Davison, M.~Rubin and G.~P.~Salam,
  arXiv:0802.2470 [hep-ph].


\bibitem{zprime_rev}
For a recent review of $Z'$ couplings, see   P.~Langacker,
  arXiv:0801.1345 [hep-ph].

\bibitem{sphericity}
  G.~Hanson {\it et al.},
  Phys.\ Rev.\ Lett.\  {\bf 35}, 1609 (1975).


\bibitem{chris}
Chris Tully, private communication.

\bibitem{punch-through}
  I.~I.~Belotelov {\it et al.},
  Phys.\ Part.\ Nucl.\ Lett.\  {\bf 4}, 343 (2007).

\bibitem{b-tag-hi-pt}
L. March, E. Ros and B. Salvachua, 
ATL-PHYS-PUB-2006-002.


\bibitem{Sudakov}
  V.~V.~Sudakov,
  Sov.\ Phys.\ JETP {\bf 3} (1956) 65
  [Zh.\ Eksp.\ Teor.\ Fiz.\  {\bf 30} (1956) 87].

\bibitem{Kaplan:2008ie}
  D.~E.~Kaplan, K.~Rehermann, M.~D.~Schwartz and B.~Tweedie,
  arXiv:0806.0848 [hep-ph].


\end{thebibliography}
\end{document}